\documentclass[aps, superscriptaddress, twocolumn, longbibliography, superscriptaddress, nofootinbib]{revtex4-1}
\usepackage[pdftex]{graphicx, color}
\usepackage{amsmath, amssymb, amscd, latexsym, bm, braket, comment, slashed}
\usepackage[colorlinks=true,pdfstartview=FitV,linkcolor=blue,citecolor=magenta,urlcolor=blue,bookmarks=true,bookmarksnumbered=true]{hyperref}

\newcommand{\appropto}{\mathrel{\vcenter{
  \offinterlineskip\halign{\hfil$##$\cr
    \propto\cr\noalign{\kern2pt}\sim\cr\noalign{\kern-2pt}}}}}

\begin{document}
\title{Kramers-Kr\"onig approach to the electric permittivity of the vacuum in a strong constant electric field}

\author{Hidetoshi Taya}
\email{hidetoshi.taya@riken.jp}
\address{RIKEN iTHEMS, RIKEN, Wako 351-0198, Japan}

\author{Charlie Ironside}
\email{charlie.ironside@curtin.edu.au}
\address{School of Electrical Engineering, Computing and Mathematical Sciences, Curtin University, Kent Street, Bentley Western Australia 6102}

\date{\today}

\begin{abstract}
We revisit the change of the electric permittivity of the QED vacuum by a strong constant electric field, motivated by the analogy between the dynamically-assisted Schwinger effect in strong-field QED and the Franz-Keldysh effect in semiconductor physics.  We develop a linear-response theory based on the non-equilibrium in-in formalism and the Furry-picture perturbation theory.  Applying the developed formalism and also utilizing the Kramers-Kr\"onig relation, we calculate the electric permittivity for wide values of the field strength and the probe frequency, including the supercritical field and/or high probe-frequency that the previous research has not fully covered.  We discover that the electric permittivity exhibits a characteristic oscillating feature in the high probe-frequency regime, which directly reflects the change of the QED-vacuum structure by the strong field.  We also establish a quantitative correspondence between the electric permittivity and the number of electron-positron pairs produced by the dynamically-assisted Schwinger effect.  
\end{abstract}

\maketitle

\section{Introduction}

At the quantum level, Dirac predicted that the vacuum of quantum electrodynamics (QED) is not just empty space, and has a structure similar to semi-conductor, called the Dirac sea~\cite{Dirac:1930ek}.  The physics of QED, thus, has some striking analogies with semiconductor due to the existence of the Dirac sea.  Semiconductor exhibits non-trivial responses when exposed to an external field because of the interaction between the field and the particles in the valence band.  Similar responses also appear in QED under a strong field, where the Dirac sea plays the role of the valence band of semiconductor.  The study of such strong-field-QED phenomena was pioneered by Heisenberg and Euler in 1936~\cite{Heisenberg:1936nmg} and has been developed by many researchers (see Refs.~\cite{DiPiazza:2011tq, Fedotov:2022ely, Hattori:2023egw} for review).  Strong-field QED attracts more attention owing to the recent availability of strong electromagnetic fields with high-power lasers (e.g., Extreme Light Infrastructure~\cite{ELI}) as well as their appearance in various physical systems under extreme conditions such as heavy-ion collisions~\cite{Bzdak:2011yy, Deng:2012pc, Hattori:2016emy, taya2023} and compact stars~\cite{Kaspi:2017fwg}.  

Various strong-field QED phenomena have been predicted theoretically.  Among those, the (Sauter-)Schwinger effect~\cite{Sauter:1931zz, Schwinger:1951nm} is one of the most intriguing and is important for our motivation to {\it revisit} the problem of the vacuum electric permittivity, which was first studied in 1950s~\cite{Toll:1952rq}.  The Schwinger effect is an analog of the dielectric breakdown of semiconductor, or the Landau-Zener(-St\"{u}ckelberg-Majorana) effect~\cite{landau1937theorie, 1932RSPSA.137..696Z, stuckelberg1933theorie, 1932NCim....9...43M}.  It states that the QED vacuum decays spontaneously against pair production under a strong constant electric field $eE \gtrsim eE_{\rm cr} := m^2$, where $m$ is the mass of a charged particle of interest, e.g., $m = 511\;{\rm keV}$ and $E_{\rm cr} = 1.32 \times 10^{18}\;{\rm V}/{\rm m}$ for electron.  Observing the Schwinger effect, or the vacuum pair production in general, is one of the biggest goals of strong-field QED experiments.  Unfortunately, the required field strength $eE_{\rm cr}$ surpasses the available field strength by several orders of magnitude~\cite{Yoon:21}.  Thus, observing the vacuum pair production is an extremely challenging task at the moment.  Nonetheless, it is proposed that the vacuum pair production can be stimulated considerably and may be within the reach even with the current and near-future lasers by superimposing a time-dependent electric field onto the constant one.  This is the idea of the dynamically-assisted Schwinger effect, proposed by Sch\"{u}tzhold, Gies, and Dunne in 2008~\cite{Schutzhold:2008pz, Dunne:2009gi}.  Making a deeper understanding of the dynamically-assisted Schwinger effect is thus an important subject of strong-field QED.

The dynamically-assisted Schwinger effect can be understood as an analog of the Franz-Keldysh effect in semiconductor physics~\cite{1958ZNatA..13..484F, keldysh1958effect, Taya:2018eng, Torgrimsson:2018xdf}.  This is an effect that stimulates the dielectric breakdown of semiconductor with the same superimposed field setup.  Conversely, it is also commonly interpreted as an effect to enhance the absorption rate of the injected time-dependent field with frequencies below the threshold band-gap energy by the presence of the strong constant electric field.  The Franz-Keldysh effect also predicts a characteristic oscillating dependence in the injected probe frequency above the threshold~\cite{PhysRev.130.549, PhysRev.130.2204}, which has been confirmed recently with strong-field-QED calculations~\cite{Taya:2018eng, Huang:2019uhf, Taya:2020pkm, Taya:2020bcd}.  Those characteristic features of the Franz-Keldysh effect below and above the threshold are originating from the change of the valence-band-electron distribution by the strong field (see, e.g., Ref.~\cite{RevModPhys.90.021002}).

It is remarkable that the Franz-Keldysh effect was predicted fifty years before the dynamically-assisted Schwinger effect.  The Franz-Keldysh effect has also been observed experimentally with various solid-state materials, since the first observation by B\"{o}er et al. with bulk semiconductors in 1958~\cite{Boer1958, Boer1959}, and has been applied to design electric devices such as electro-absorption modulator that are widely used in high-speed digital communications in our daily lives.  The Franz-Keldysh effect is, thus, a well-known and -tested phenomenon in semiconductor physics.

Therefore, it should be instructive to import the wisdom of semiconductor physics to strong-field QED.  As such, we in this paper revisit the change of the electric permittivity of the QED vacuum by a strong constant electric field.  This is analogous to that of semiconductor investigated extensively in the context of {\it electroreflectance}: Suppose we have a semiconductor and inject a probe electric field onto it.  The electric response of the material to the probe can be characterized by the so-called electric permittivity.  In general, the electric permittivity (in the frequency space) takes complex values.  The real and imaginary parts, respectively, account for the dispersive and absorptive phenomena due to the interaction between the valence-band electrons and the injected probe.  The material property changes with environments such as the presence of an external field, so does the electric permittivity.  The Franz-Keldysh effect describes the change of the absorption rate of the probe under a constant electric field, which is directly related to the imaginary part of the electric permittivity.  The imaginary part, thus, exhibits characteristic features originating from the Franz-Keldysh effect.  The change of the imaginary part in turn implies that the real part should also be modified, as the imaginary and real parts are not independent quantities but are related with each other through the Kramers-Kr\"onig relation because of causality~\cite{deL.Kronig:26,1571135649980336384, Toll:1956cya}.  Meanwhile, the electric permittivity is closely related to the refractive index of the material (e.g., for non-magnetic materials, the refractive index is simply given by the square of the electric permittivity).  Therefore, the change of the electric permittivity means the refractive index is no longer constant and is modified significantly by a constant electric field, which is called electroreflectance and has been discussed widely in the semiconductor community~\cite{PhysRevLett.14.138, PhysRev.139.A560, PhysRev.143.564, PhysRev.145.628, PhysRev.153.972, Hutchings1992, 10.1088/978-1-6817-4521-3}.

Considering the analogy between QED and semiconductor and also that between the dynamically-assisted Schwinger effect and the Franz-Keldysh effect, it is natural to expect a similar modification to the electric permittivity of the QED vacuum should happen and can be analyzed in an analogous manner to semiconductor physics.  The main purpose of this paper is to explore this idea.  A common strategy in semiconductor physics to calculate the electric permittivity is first to calculate the absorption rate of the probe, which can be converted into the imaginary part, and then to utilize the Kramers-Kr\"onig relation to get the real part from the imaginary part~\cite{PhysRevLett.14.138, PhysRev.139.A560, PhysRev.143.564, PhysRev.145.628, PhysRev.153.972, Hutchings1992, 10.1088/978-1-6817-4521-3} (cf. a similar Kramers-Kr\"onig approach has recently been adopted in strong-field QED, in connection to the non-linear Breit-Wheeler process, in Refs.~\cite{Heinzl:2006pn, Borysov:2022cwc, King:2023eeo}).  We shall provide a theoretical foundation of this semiconductor approach in QED based on the non-equilibrium in-in formalism of quantum-field theory (see, e.g., Refs.~\cite{Schwinger:1960qe, kadanoff1962quantum, Keldysh:1964ud}) and the Furry-picture perturbation theory (see, e.g., Refs.~\cite{Furry:1937zz, Fradkin:1991zq}).  We shall, then, use the developed formalism to explicitly calculate the electric permittivity of the QED vacuum and discuss its quantitative features, including similarities/differences with semiconductor physics.

We have several other motivations, besides the analogy to semiconductor physics: (i) In the presence of an electric field, the system becomes genuinely non-equilibrium.  Typically, what is discussed in the literature are the cases of, e.g., magnetic fields, constant-crossed fields, and the weak-field limit.  In such situations, there is no energy supply from the fields and therefore the system stays equilibrium, i.e., the situation is totally different from our non-equilibrium electric case.  In a non-equilibrium situation, a careful distinction between the in- and out-states must be made, and the conventional in-out formalism cannot be applied because the physical observables are in-in expectation values and not the in-out amplitudes~\cite{Copinger:2018ftr}.  To the best of our knowledge, there has been no such in-in formulation of the electric permittivity, or in general QED response functions.  

(ii) The idea of the change of the electric permittivity, or the refractive index, in strong-field QED is actually not new.  It has been pointed out by the classical work by Toll in 1952~\cite{Toll:1952rq} and since then has been under intensive investigations both theoretically (see, e.g., Ref.~\cite{King:2015tba}) and experimentally (e.g., PVLAS~\cite{Ejlli:2020yhk}).  Nonetheless, the previous studies focus mainly on the equilibrium situations and also are based typically on weak-field- and low-frequency- approximations such as the Heisenberg-Euler approach and semi-classical methods.  Thus, little is known for {\it electric} field and {\it beyond} the weak-field and low-frequency limit (cf. Ref.~\cite{Karbstein:2013ufa} for the refractive index in a very strong electric-field limit based on the electromagnetic duality).  Our calculation does not assume such weak-field and low-frequency conditions, and therefore is able to reveal novel aspects of the electric permittivity in a new parameter regime.  

(iii) As far as the authors are aware, there is no clear argument on the relation between the dynamically-assisted Schwinger effect and the electric permittivity.  In semiconductor physics, it is well known that the imaginary part of the electric permittivity is directly related to the dielectric energy loss of material, which can then be related to the number of pairs produced via the Franz-Keldysh effect.  Establishing a similar relation in QED is useful, in particular, for experimental purpose.  Namely, using the relation, the observation of the pair production can be used to directly quantify the response function of the QED vacuum, i.e., the imaginary part of the electric permittivity and in turn the real part through the Kramers-Kr\"{o}nig relation.  Conversely, the measurement of either of the imaginary or real part can be used to confirm the dynamically-assisted Schwinger effect.  Note that a similar idea has been advocated in Refs.~\cite{Borysov:2022cwc, King:2023eeo} in the context of the non-linear Breit-Wheeler process.  We remark that what is relevant for the number of pairs in the dynamically-assisted Schwinger effect is the in-in expectation value, not the in-out amplitudes, and therefore arguments based on the in-out formalism, such as the optical theorem, are not very appropriate to establish the relation.  Therefore, a careful in-in formulation (i) is crucial to achieve the motivation (iii).

The organization of this paper is as follows.  We first clarify our physical setup and assumptions in Sec.~\ref{sec:2}.  We then develop a linear-response theory to calculate the electric permittivity based on the non-equilibrium in-in formalism of quantum-field theory and the perturbation theory in the Furry picture in Sec.~\ref{sec:3}.  Using this, we explicitly calculate the electric permittivity and quantitatively discuss its properties in Sec.~\ref{sec:4}.  We summarize our results and discuss possible experimental implications in Sec.~\ref{sec:5}.  We also have two Appendices~\ref{AppendixA} and \ref{app:C}, where some technical details of the derivations of analytical formulas are explained.

{\it Notation and convention}.  We adopt the natural units $\hbar = c = 1$ and $\epsilon_0 = \mu_0 = 1$, where $\hbar$ is the Planck constant and $c$, $\epsilon_0$, and $\mu_0$ are, respectively, the speed of light, electric permittivity, and magnetic permeability of the vacuum without external fields.  Our metric convention is the mostly minus $g_{\mu\nu} := {\rm diag}(+1,-1,-1,-1)$.  We express the components of the coordinate four vector $x^\mu$ as $x^\mu =: (t,{\bm x}) =: (t,x,y,z)$.  We occasionally use the Roman letter to express the absolute value of a three vector as $X := |{\bm X}|$.  We sum over repeated indices implicitly, otherwise stated.

\section{Setup} \label{sec:2}

We consider QED in the presence of a strong electric field and wish to understand the response, i.e., electric permittivity, against a probe electric wave.  The electron\footnote{Our formulation is equally valid for charged particles obeying the Dirac equation (\ref{eq:1}).  This means that the ``electron'' here can actually be any charged elementary/composite/quasi fermion such as quark and proton (so long as the external field is not too strong that the point-like-particle treatment is justified).  Only the difference is that the value of the mass $m$ changes if we change the particle species.  Therefore, we do not specify a specific value of the mass $m$ nor take it to be the electron one, unless necessary.  } dynamics is determined by the Dirac equation,
\begin{align}
	0 = \left[ {\rm i}\slashed{\partial} - e\slashed{A} + m \right] \psi \;, \label{eq:1} 
\end{align}
where $\psi$ is the electron field operator and $A^\mu$ is the gauge potential for the total electric field, i.e., the sum of the strong and probe electric fields.  

We assume that the strong field is constant in spacetime and that the probe field is spatially homogeneous but is dependent on time with a monochromatic frequency $\omega >0$ (which can be interpreted as a sort of time-like off-shell photons).  Namely,  
\begin{align}
	{\bm A}(t) = \underbrace{ - \bar{E} t {\bf e}_z }_{=: \bar{\bm A}(t) } + \underbrace{ \frac{{\bm {\mathcal E}}}{\omega} \cos (\omega t+\phi) }_{=: {\bm {\mathcal A}}(t)} \; , \label{eq:2}
\end{align} 
where we fixed the gauge as $A^\mu = (0, {\bm A})$.  The gauge potentials for the strong and probe electric fields are denoted by $\bar{\bm A}$ and ${\bm {\mathcal A}}$, respectively.  We took the direction of the strong electric field $\bar{\bm E} := -\partial_t \bar{\bm A}$ along the $z$-axis ${\bf e}_z := \bar{\bm E}/\bar{E} $.

We make several more assumptions for the electric fields for simplicity.  First, we assume that the probe field is very weak, so as to justify a perturbative treatment in $|{\bm {\mathcal E}}|$.  Second, we assume formally that the electric fields are switched-off adiabatically at the infinite past and future $\bar{\bm A}, {\bm {\mathcal A}} \to {\rm const}.$, so that the asymptotic states are well-defined.  Third, we neglect back-reaction from electrons onto the fields; otherwise, the electric field changes its value in time, for which the electric permittivity is not well-defined.

Although we focus on the simple field configuration (\ref{eq:2}), it is in principle straightforward to extend our formalism to general field configurations such as strong fields with magnetic components and a more realistic on-shell photon probe.  This shall be discussed in future publication.  The reason why we consider the configuration (\ref{eq:2}) in this paper is that it is the simplest and is widely used in semiconductor physics.  Therefore, it provides a good starting point to pursue the analogy between QED and semiconductor physics.  In fact, the electric permittivity can be measured with dielectric spectroscopy (see, e.g., Ref.~\cite{kremer2002broadband}), whose typical setup corresponds to Eq.~(\ref{eq:2}): A material to be investigated is sandwiched by a set of plates, which form up a capacitor, and an alternating current (AC) voltage is applied to the capacitor.  Then, the measurement of the impedance of the capacitor can determine the electric permittivity of the material.  In our case, the material corresponds to the vacuum in a strong electric field $\bar{\bm E}$, and the AC voltage plays the role of the probe ${\bm {\mathcal E}}$.

\section{Theory} \label{sec:3}

We calculate the change of the electric permittivity for the probe electric wave (\ref{eq:2}) in the presence and absence of the strong constant electric field $\bar{\bm E}$.  The calculation of the electric permittivity is reduced to evaluating the in-in vacuum expectation value of the electric current (see Sec.~\ref{sec:3A}).  We first write down a formal expression for the electric current, and thereby the electric permittivity, by developing a linear-response theory based on the in-in formalism of quantum-field theory in the Furry picture, where the probe field is treated perturbatively while the strong field non-perturbatively (see Sec.~\ref{sec:3B}).  The central task shall be then how to evaluate the formal expression, which has two difficulties.  The first difficulty is that evaluation of the real part of the electric permittivity (corresponding to evaluating a fermion one-loop diagram) in general suffers from subtleties of quantum-field theory such as the ultraviolet divergence and therefore is complicated.  In contrast, the imaginary part (corresponding to a tree diagram, obtained by cutting the fermion loop diagram) does not have such complications, and the evaluation is far easier.  We shall thus avoid directly evaluating the electric permittivity; instead, we shall calculate the imaginary part first and then make use of the Kramers-Kr\"{o}nig relation to get the real part, which is much easier than the direct evaluation.  This is actually a widely-adopted approach in the semiconductor context~\cite{PhysRevLett.14.138, PhysRev.139.A560, PhysRev.143.564, PhysRev.145.628, PhysRev.153.972, Hutchings1992, 10.1088/978-1-6817-4521-3}, and therefore fits our purpose of pursuing the QED analog of semiconductor physics.  In Sec.~\ref{sec:3C}, we shall justify this treatment based on our in-in formulation, which is manifestly causal and naturally yields the Kramers-Kr\"{o}nig relation.  The second difficulty is that there appears infrared divergence in the low-frequency limit of the probe $\omega \to 0$, implying the need of resummation of higher-order processes in the probe (see Sec.~\ref{sec:3D}).  We here deal with the infrared divergence rather phenomenologically, without explicitly carrying out the resummation, by relating the electric permittivity to the number of pairs produced by the dynamically-assisted Schwinger effect (see Sec.~\ref{sec:3E}) and determining a physically-reasonable counter term by requiring the number of pairs to be finite (see Sec.~\ref{sec:3F}).  Our final formula for the change of the electric permittivity shall be summarized in Sec.~\ref{sec:3G}.

We shall be dedicated to the theoretical foundation only in this section and put all the quantitative and phenomenological discussions in Sec.~\ref{sec:4}.

\subsection{Definition and generality of the electric permittivity} \label{sec:3A}

We begin with clarifying the definition and the underlying assumptions of electric permittivity.  To define the electric permittivity, we first need to define the electric displacement vector.  To this end, we write down the Amp\`{e}re-Maxwell law for our spatially homogeneous configuration (\ref{eq:2}): 
\begin{align}
	\dot{\bm E} = -{\bm J}_{\rm ext} - {\bm J}_{\rm vac} \;,
\end{align}
where ${\bm E} = \bar{\bm E} + {\bm {\mathcal E}}$ is the total electric field, ${\bm J}_{\rm ext}$ is the external current, and 
\begin{align}
	{\bm J}_{\rm vac} := e \braket{ {\rm vac}; {\rm in} | \bar{\psi} {\bm \gamma} \psi | {\rm vac}; {\rm in} }  \label{v2er2:4}
\end{align}
is the contribution from the vacuum polarization.  The state $\ket{{\rm vac}; {\rm in}}$ is the initial vacuum, and $\gamma^\mu := (\gamma^0, {\bm \gamma}) := (\gamma^0, \gamma^1, \gamma^2, \gamma^3)$ are the Dirac matrices (the precise definition of the initial vacuum shall be clarified below in Sec.~\ref{sec:3B}).  We stress that the in- and out-states must be distinguished carefully.  Indeed, the electric field brings time dependence to the system by supplying energy, and hence the system is not in equilibrium.  The electric current (or physical observables in general) is an in-in expectation value, which is distinct from in-out amplitudes (which is not direct physical observables) that are typically calculated in the literature.

We can naturally introduce the total electric displacement vector ${\bm D}$ as
\begin{align}
	\dot{\bm D} := -{\bm J}_{\rm ext} \;,  \label{gqerqed:6}
\end{align}
i.e., ${\bm D}$ is defined as a renormalized electric field such that it absorbs the vacuum-polarization effect:  
\begin{align}
	{\bm D}(t) = {\bm E}(t) + \int^t_{-\infty} {\rm d}t'\,{\bm J}_{\rm vac}(t') \;.  \label{eqwrwqdqw:381}
\end{align}
We have implicitly imposed an initial condition ${\bm D} = {\bm E}$ at $t = -\infty$, as we are physically interested in a situation where the electric field is (adiabatically) applied at $t=-\infty$ and then the non-trivial vacuum polarization develops from null.

As we are interested in the electric permittivity for the probe, we need to decompose the total electric displacement vector ${\bm D}$ (\ref{eqwrwqdqw:381}) into the strong $\bar{\bm D}$ and probe ${\bm {\mathcal D}}$ parts.  To do this, we expand the vacuum polarization ${\bm J}_{\rm vac}$ in the weak probe field ${\bm {\mathcal E}}$, keeping the strong field $\bar{\bm E}$ exactly, as 
\begin{align}
	{\bm J}_{\rm vac}(t; \bar{\bm E}, {\bm {\mathcal E}}) 
	&=: {\bm J}_{\rm vac}(t; \bar{\bm E}, {\bm 0}) \label{g23r23r:34} \\
	&\quad + \int^{+\infty}_{-\infty} {\rm d}t'\,\sigma_{ij}\left(t,t' ; \bar{\bm E} \right) {\mathcal E}_j(t') + {\mathcal O}(|{\bm {\mathcal E}}|^2) \;. \nonumber
\end{align}
To be precise, the perturbative expansion in ${\bm {\mathcal E}}$ needs a special care in the low-frequency limit $\omega \to 0$, where non-linear corrections by ${\bm {\mathcal E}}$ can become non-negligible even though the strength $|{\bm {\mathcal E}}|$ is weak and can lead to divergent behaviors.  We shall come back to this point in Sec.~\ref{sec:3F}.  We also remark that the function $\sigma_{ij}$ depends on the two time variables in general; only when the system is invariant under the time translation, can $\sigma_{ij}$ become a single-variable function.  This is important for the definition of the electric permittivity to make sense, which we shall discuss in more detail below.

We can now naturally identify the electric displacement vector for the probe ${\bm {\mathcal D}}$.  Substituting Eq.~(\ref{g23r23r:34}) into Eq.~(\ref{eqwrwqdqw:381}), picking up the probe-dependent terms, and dropping the non-linear term ${\mathcal O}(|{\bm {\mathcal E}}|^2)$, we find
\begin{align}
	{\mathcal D}_i(t) = {\mathcal E}_i(t) +  \int^t_{-\infty} {\rm d}t'\, \int^{+\infty}_{-\infty} {\rm d}t''\,\sigma_{ij}\left(t', t'' ; \bar{\bm E} \right) {\mathcal E}_j(t'') \;. \label{eqwrwqdqw:3as81}
\end{align}
The zeroth-order term ${\bm J}_{\rm vac}(t; \bar{\bm E}, {\bm 0})$ is probe independent, and hence contributes only to the electric displacement vector for the strong field $\bar{\bm D}$ and is irrelevant to ${\bm {\mathcal D}}$.

We are in the position to define the electric permittivity for the probe $\epsilon_{ij}$.  It is standard to identify the electric permittivity as a linear coefficient between the applied probe and the corresponding displacement vector in the Fourier space (see textbooks on electromagnetism in matter, e.g., Ref.~\cite{lan84}).  Namely, 
\begin{align}
	\tilde{\mathcal D}_i(\omega)
	=: \epsilon_{ij}(\omega) \tilde{\mathcal E}_j(\omega)  \label{fqereqr;q} \;,
\end{align}
where $\tilde{f}(\omega) := \int^{+\infty}_{-\infty}{\rm d}t\,{\rm e}^{+{\rm i}\omega t} f(t)$ is the Fourier transformation of a single-variable function $f$.  Note that, in more general, the electric permittivity is defined as a function of the wave four-vector $k^\mu = (\omega, {\bm k})$, with ${\bm k}$ being the wave three-vector of the probe.  In our setup (\ref{eq:2}), the probe field is assumed to be spatially homogeneous and therefore the value of ${\bm k}$ is fixed to be ${\bm k} = {\bm 0}$.

We wish to apply the definition (\ref{fqereqr;q}) to our problem, but this needs some discussion.  Namely, the definition (\ref{fqereqr;q}) is not applicable for general non-equilibrium situations, where $\tilde{\bm {\mathcal D}}$ and $\tilde{\bm {\mathcal E}}$ are related with each other by a convolution integral in the frequency $\omega$, instead of the simple product form (\ref{fqereqr;q}).  Indeed, sending Eq.~(\ref{eqwrwqdqw:3as81}) to the Fourier space yields
\begin{align}
	\tilde{\mathcal D}_i(\omega)
	&=  \int^{+\infty}_{-\infty} \frac{{\rm d}\omega'}{2\pi}\left[ 2\pi \delta(\omega-\omega')   +  \frac{{\rm i}}{\omega} \tilde{\sigma}_{ij}\left( \omega, \omega' \right) \right] \tilde{\mathcal E}_j(\omega') \;, \label{fqerq:11}
\end{align}
where the Fourier transformation of the two-variable function $\tilde{\sigma}_{ij}$ is defined similarly to the single-variable case as $\tilde{f}(\omega,\omega') := \int^{+\infty}_{-\infty}{\rm d}t{\rm d}t'\,{\rm e}^{+{\rm i}\omega t} {\rm e}^{+{\rm i}\omega' t'} f(t,t')$.  From Eq.~(\ref{fqerq:11}), it is evident that the standard definition (\ref{fqereqr;q}) is applicable only when $\tilde{\sigma}_{ij}$ is sharply peaked at $\omega = \omega'$ and can be decomposed with a delta function as
\begin{align}
	\tilde{\sigma}_{ij}\left( \omega, \omega' \right) 
	=: 2\pi \delta(\omega-\omega') \times \tilde{\sigma}_{ij}\left( \omega \right) \;, \label{rqweqasdwe:12}
\end{align}
for which case the electric permittivity is given by
\begin{align}
	\epsilon_{ij}(\omega) 
	=  1   +   \frac{{\rm i}}{\omega} \tilde{\sigma}_{ij}\left( \omega \right) \;. \label{gqerqeqewq:13}
\end{align}
The condition (\ref{rqweqasdwe:12}) is satisfied if and only if $\sigma_{ij}$ in the coordinate space depends solely on the difference between the two variables $t-t'$ and is independent of the sum $t+t'$, i.e., $\sigma_{ij}$ is invariant under the time translation.  It is achieved only if the system is in equilibrium or in a steady state.  For general strong fields, in particular under strong time-dependent electric fields, the system is genuinely in non-equilibrium and the condition (\ref{rqweqasdwe:12}) is not satisfied, meaning that the standard definition of electric permittivity (\ref{fqereqr;q}) does not make sense.

Even though the standard definition (\ref{fqereqr;q}) is invalid for general electromagnetic fields, this is, fortunately, not the case in our problem.  That is, a steady state is realized for a {\it constant} electric field $\bar{\bm E}$ and hence $\sigma_{ij}$ can be reduced to a single-variable function.  Therefore, the definition (\ref{fqereqr;q}), or Eq.~(\ref{gqerqeqewq:13}), can be applied.  This seems not possible to be shown by general arguments but requires explicit calculations, which we shall do in the next section \ref{sec:3B}.

We are thus able to calculate the electric permittivity in our problem according to the standard definition (\ref{gqerqeqewq:13}).  Before proceeding to the detailed calculations, let us, for later use, make several remarks on the general properties of the electric permittivity.  

The first remark is that in general the electric permittivity $\epsilon_{ij}$ can take non-zero values even in the absence of the strong field $\bar{\bm E}$.  This is because the probe alone (i.e., even without $\bar{\bm E}$) can excite electron-positron pairs from the vacuum and thus can contribute to the vacuum polarization ${\bm J}_{\rm vac}$, once the frequency gets above the threshold band-gap energy $\omega > 2m$.  Therefore, to see the strong-field effects in a clearer manner, it is more useful to look into the difference between the presence and absence of the strong field.  Thus, in what follows, we shall focus on the difference, 
\begin{align}
	\Delta \epsilon_{ij}(\omega; \bar{\bm E}) 
	:=  \epsilon_{ij}(\omega; \bar{\bm E} \neq {\bm 0} )  -  \epsilon_{ij}(\omega; \bar{\bm E} = {\bm 0}) \;, \label{qefqedqas:14}
\end{align}
instead of the original value of the electric permittivity $\epsilon_{ij}$.  This is also advantageous to our Kramers-Kr\"{o}nig approach, as the high-frequency behavior of $\Delta \epsilon_{ij}$ is better than $\epsilon_{ij}$, resulting in better numerical convergence when performing the Hilbert transformation in the Kramers-Kr\"{o}nig relation.  

The second remark is that the tensor structure of the electric permittivity can be constrained by symmetry.  Our system (\ref{eq:2}) has no special directions other than ${\bf e}_z$ set by the strong electric field $\bar{\bm E}$.  Accordingly, the physics should not change by rotating the system around ${\bf e}_z$ (rotation symmetry) and by reflecting it with respect to ${\bf e}_z$ (parity symmetry).  This means that the electric permittivity should be invariant under the corresponding transformations: 
\begin{align}
	\epsilon = R \epsilon R^\dagger \;, \label{sfqweqw:16}
\end{align}
where
\begin{align}
	R = \begin{pmatrix} \cos \theta & -\sin\theta & 0 \\ \sin \theta & \cos\theta & 0 \\ 0 & 0& 1 \end{pmatrix} \ {\rm and}\  \begin{pmatrix} \pm 1 & 0 & 0 \\ 0 & \pm 1 & 0 \\ 0 & 0 & 1 \end{pmatrix} \;,
\end{align}
for the rotation with an angle $\theta$ and the parity transformation, respectively.  Solving Eq.~(\ref{sfqweqw:16}), we find that the electric permittivity must be diagonal and that the transverse components take the same value, $\epsilon_{xx} = \epsilon_{yy} =: \epsilon_\perp$, as
\begin{align}
	\epsilon_{ij} = \begin{pmatrix} \epsilon_\perp & 0 & 0 \\ 0 & \epsilon_\perp & 0 \\ 0 & 0 & \epsilon_\parallel \end{pmatrix} \;.  \label{qweqw:17}
\end{align}
In general, $\epsilon_\perp \neq \epsilon_{zz} =: \epsilon_\parallel$, i.e., the vacuum is birefringent.  This is because we have the special direction $\bar{\bm E} \propto {\bf e}_z$ and the response should distinguish ${\bm {\mathcal E}} \parallel {\bf e}_z$ and ${\bm {\mathcal E}} \perp {\bf e}_z$.  Note that for general field configurations, the symmetry structure can be changed and accordingly the electric permittivity can have a different tensor structure; e.g., in a strong magnetic field, the off-diagonal elements are not necessarily vanishing, as the parity symmetry is broken.  

The final remark is that the real and imaginary parts of the electric permittivity must be even and odd functions in $\omega$:
\begin{align}
	{\rm Re}\,\epsilon(\omega) = {\rm Re}\,\epsilon(-\omega) \ \ {\rm and}\ \ 
	{\rm Im}\,\epsilon(\omega) = -{\rm Im}\,\epsilon(-\omega) \;. \label{egqefsreqr:18}
\end{align}
This follows from the reality of the probe and the electric displacement vector in the coordinate space ${\bm {\mathcal X}}(t) \in {\mathbb R}^3$ (${\bm {\mathcal X}} = {\bm {\mathcal E}}, {\bm {\mathcal D}}$).  From reality, we find that the Fourier transformations must satisfy $\tilde{\bm {\mathcal X}}(\omega) = [\tilde{\bm {\mathcal X}}(-\omega)]^*$, substituting which into the definition of $\epsilon_{ij}$ (\ref{fqereqr;q}) immediately yields the relation (\ref{egqefsreqr:18}).

\subsection{Linear-response theory in a strong electric field}  \label{sec:3B}

We calculate the leading-order correction to the current $\sigma_{ij}$ (\ref{g23r23r:34}) to obtain a formal expression for the electric permittivity via Eq.~(\ref{gqerqeqewq:13}), by developing a linear-response theory in strong fields based on the in-in formalism of quantum-field theory and the perturbation theory in the Furry picture.

To calculate $\sigma_{ij}$ (\ref{g23r23r:34}), we need to expand the current ${\bm J}_{\rm vac}$, or the electron field operator $\psi$ in the expectation value (\ref{v2er2:4}), with respect to the probe field ${\bm {\mathcal E}}$.  Using the Green function technique, it is straightforward to solve the Dirac equation (\ref{eq:1}) perturbatively in ${\bm {\mathcal E}}$ as
\begin{align}
	\psi(t,{\bm x}) 
	&= \psi^{\rm in}(t,{\bm x}) \nonumber\\
	&\quad + e\int{\rm d}^4x'\, S^{\rm R}(t,t'; {\bm x},{\bm x}') \slashed{\mathcal A}(t') \psi^{\rm in}(t',{\bm x}')   \nonumber\\
	&\quad +   {\mathcal O}(|{\bm {\mathcal E}}|^2) \;. \label{fqewrqe:435}
\end{align}
We have imposed the boundary condition $\psi^{\rm in} := \lim_{t \to -\infty} \psi$, with $\psi^{\rm in}$ being the zeroth-order solution to the Dirac equation only with the strong field $\bar{\bm E}$, 
\begin{align}
	\left[ {\rm i}\slashed{\partial} - e\bar{\slashed{A}} - m \right] \psi^{\rm in} = 0 \;,  \label{gfqewqwd:19}
\end{align}
because the initial condition of our problem is ${\bm {\mathcal E}} \to 0$ at $t \to -\infty$.  To achieve the boundary condition, the Green function $S^{\rm R}$ must be the retarded one such that 
\begin{align}
	\left\{ \begin{array}{l} \displaystyle \left[ {\rm i}\slashed{\partial} - e\bar{\slashed{A}} - m \right] S^{\rm R}(t,t';{\bm x},{\bm x}') = \delta(t-t')\delta^3({\bm x}-{\bm x}') \;, \\[12pt] S^{\rm R}(t,t';{\bm x},{\bm x}') = 0 \ \ {\rm for}\ t-t'<0 \;. \end{array} \right.  \label{eraqerq:16}
\end{align}

To move to quantum-field theory, we need to quantize the (asymptotic) electron-field operator $\psi^{\rm in}$.  To do so, we first expand $\psi^{\rm in}$ with the mode functions $\psi^{\rm in}_{\pm,{\bm p},s}$ as 
\begin{align}
	\psi^{\rm in}(t,{\bm x}) 
	&= \sum_{s = \pm 1} \int {\rm d}^3{\bm p} \frac{{\rm e}^{+{\rm i}{\bm p}\cdot {\bm x}}}{(2\pi)^{3/2}} \nonumber\\
	&\quad \times \left[  \psi^{\rm in}_{+,{\bm p},s}(t) a^{{\rm in}}_{{\bm p},s}   +   \psi^{\rm in}_{-,{\bm p},s}(t) b^{{\rm in}\dagger}_{-{\bm p},-s}  \right] \;, \label{fqewqw:21}
\end{align}
with ${\bm p}$ and $s$ specifying (canonical) momentum and spin, respectively.  We impose a plane-wave boundary condition onto the mode functions, 
\begin{align}
	&\lim_{t \to -\infty} \begin{pmatrix} \psi^{\rm in}_{+,{\bm p},s} \\ \psi^{\rm in}_{-,{\bm p},s} \end{pmatrix} \nonumber\\
	&= \begin{pmatrix} u_{{\bm p}-e\bar{\bm A}(-\infty),s} \\ v_{-{\bm p}-e\bar{\bm A}(-\infty),s} \end{pmatrix} {\rm e}^{\mp {\rm i}\sqrt{m^2 + ({\bm p}-e\bar{\bm A}(-\infty))^2}t} \;,   \label{eq-10}
\end{align}
where $u_{{\bm p},s}$ and $v_{{\bm p},s}$ are the usual Dirac spinors with the normalization $u_{{\bm p},s}^\dagger u_{{\bm p},s'} = v_{{\bm p},s}^\dagger v_{{\bm p},s'} = \delta_{s,s'}$ and $u_{{\bm p},s}^\dagger v_{{\bm p},s'} = 0$.  Note that the normalization of the Dirac spinors automatically normalizes the mode function as $\psi^{{\rm in}\dagger}_{\pm,{\bm p},s} \psi^{{\rm in}}_{\pm',{\bm p},s'} = \delta_{\pm,\pm'} \delta_{s,s'}$.  Then, imposing the canonical commutation relation onto $\psi^{\rm in}$, $a_{{\bm p},s}^{\rm in}$ and $b_{{\bm p},s}^{\rm in}$ are quantized and satisfy the standard anti-commutation relations, 
\begin{align}
	&\delta_{s,s'} \delta^3({\bm p}-{\bm p}') 
	= \{ a_{{\bm p},s}^{\rm in} \; , \; a_{{\bm p}',s'}^{{\rm in}\dagger}  \}
	= \{ b_{{\bm p},s}^{\rm in} \; , \; b_{{\bm p}',s'}^{{\rm in}\dagger}  \} \nonumber\\
	&0 = ({\rm others}) \;. \label{qerqerq:19}
\end{align}
As the external electric field ${\bm E}$ goes vanishing at $t \to -\infty$, we can naturally identify the operators $a_{{\bm p},s}^{\rm in}$ and $b_{{\bm p},s}^{\rm in}$ to be the annihilation operators for an electron and a positron, respectively, with momentum ${\bm p}$ and spin $s$.  Accordingly, the initial vacuum can be identified as a state such that
\begin{align}
	0 = a_{{\bm p},s}^{\rm in} \ket{{\rm vac;in}} = b_{{\bm p},s}^{\rm in} \ket{{\rm vac;in}}  \;,
\end{align}
for any ${\bm p}$ and $s$.  The normalization of $\ket{{\rm vac;in}}$ is arbitrary, and we take $1 = \braket{{\rm vac;in} | {\rm vac;in}} $ for our convenience.

Plugging the perturbative solution (\ref{fqewrqe:435}) with the quantization (\ref{qerqerq:19}) into the in-in expectation value (\ref{v2er2:4}), we get
\begin{align}
	{\bm J}_{\rm vac}(t) 
	&= \sum_s \int \frac{{\rm d}^3{\bm p}}{(2\pi)^3} \,\bar{\psi}_{-,{\bm p},s}^{\rm in}(t) {\bm \gamma} \psi_{-,{\bm p},s}^{\rm in}(t)  \nonumber\\
		&\quad - 2e^2 \int^{+\infty}_{-\infty} {\rm d}t' \, \Theta(t-t') {\mathcal A}_\mu(t')  \nonumber\\
			&\quad\quad \times {\rm Im} \,{\rm tr} \sum_{s,s'} \int \frac{{\rm d}^3{\bm p}}{(2\pi)^3} {\bm \gamma} S^{\rm in}_{-,{\bm p},s}(t,t')  \gamma^\mu  S^{\rm in}_{+,{\bm p},s'}(t',t) \nonumber\\
		&\quad +   {\mathcal O}(|{\bm {\mathcal E}}|^2) \;, \label{afqerqrqe:21}
\end{align}
where $\Theta(t)$ is the Heaviside step function and we have used that the retarded Green function $S^{\rm R}$ can be expressed in terms of the mode functions $\psi^{\rm in}_{\pm, {\bm p},s}$ as
\begin{align}
	&S_{\rm R}(t,t';{\bm x},{\bm x}') \nonumber\\
	&= -{\rm i} \Theta(t-t') \left\{ \psi^{\rm in}(t,{\bm x})  \,,\,\bar{\psi}^{\rm in}(t,{\bm x}')  \right\}  \nonumber\\
	&= -{\rm i} \Theta(t-t') \sum_{s} \int {\rm d}^3{\bm p} \frac{{\rm e}^{+{\rm i}{\bm p}\cdot({\bm x}-{\bm x}')}}{(2\pi)^3} \sum_{\pm} \psi^{\rm in}_{\pm,{\bm p},s}(t)\bar{\psi}^{\rm in}_{\pm,{\bm p},s}(t') \nonumber\\
	&=: -{\rm i} \Theta(t-t') \sum_{s} \int {\rm d}^3{\bm p} \frac{{\rm e}^{+{\rm i}{\bm p}\cdot({\bm x}-{\bm x}')}}{(2\pi)^3} \sum_{\pm}  S^{\rm in}_{\pm,{\bm p},s}(t,t') \;.
\end{align}
We emphasize the existence of the step function in Eq.~(\ref{afqerqrqe:21}), which is mathematically originating from the retarded boundary condition (\ref{eraqerq:16}).  Physically, it is the manifestation of causality: the probe field cannot drive current before it is applied.

Comparing Eq.~(\ref{afqerqrqe:21}) with the definition of $\sigma_{ij}$ (\ref{g23r23r:34}), we find
\begin{align}
	\partial_{t'} \sigma_{ij}\left(t,t'\right) 
	= \Theta(\tau) \Pi_{ij}(T,\tau) 
\end{align}
where
\begin{align}
	T := \frac{t+t'}{2}\ \ {\rm and}\ \ \tau := t-t' \;,
\end{align}
and
\begin{align}
	&\Pi_{ij}(T,\tau) \label{geqrqsa:29} \\
	&:= 2e^2 \,{\rm Im}\,{\rm tr}  \sum_{s,s'} \int \frac{{\rm d}^3{\bm p}}{(2\pi)^3} \nonumber\\
	&\quad \times \gamma^i  S^{\rm in}_{-,{\bm p},s}\left( T+\frac{\tau}{2}, T-\frac{\tau}{2} \right)  \gamma^j   S^{\rm in}_{+,{\bm p},s'} \left( T-\frac{\tau}{2},T+\frac{\tau}{2} \right) \;. \nonumber
\end{align}
Note that $\Pi_{ij}$ is similar to the photon polarization tensor in the in-out calculations but has a bit different form (e.g., no Feynman propagator), since ours is an in-in calculation.

As stressed in Sec.~\ref{sec:3A}, the standard definition of the electric permittivity (\ref{fqereqr;q}) makes sense only if $\sigma_{ij}$ does not depend on $t+t' \propto T$.  This is the case for our constant electric field $\bar{\bm E}$.  To show this, we remind that the field operator $\psi^{\rm in}$ satisfies the Dirac equation (\ref{gfqewqwd:19}) and the corresponding mode equation is 
\begin{align}
	0 = \left[ {\rm i}\gamma^0 \partial_t - {\bm \gamma}_\perp \cdot {\bm p}_\perp - \gamma^3(p_z + e\bar{E}t) - m  \right] \psi^{\rm in}_{\pm,{\bm p},s} \;. \label{fqewdasx;30}
\end{align}
The key point here is that the $t$ dependence always appears together with $p_z$.  This means that the time dependence of the mode function can be absorbed to the momentum as
\begin{align}
	\psi^{\rm in}_{\pm,{\bm p},s}(t) = \psi^{\rm in}_{\pm,{\bm p} + e\bar{\bm E}t',s}(t-t') \label{tqewda:31}
\end{align}
for any $t'$.  The $T$ dependence of $\Pi_{ij}$ can then be integrated out by shifting the momentum integration variable as 
\begin{align}
	\Pi_{ij} 
	&= 2e^2 \,{\rm Im}\,{\rm tr}  \sum_{s,s'} \int \frac{{\rm d}^3{\bm P}}{(2\pi)^3} \nonumber\\
	&\quad \times \gamma^i  S^{\rm in}_{-,{\bm P},s}\left( +\frac{\tau}{2}, -\frac{\tau}{2} \right)  \gamma^j   S^{\rm in}_{+,{\bm P},s'} \left( -\frac{\tau}{2},+\frac{\tau}{2} \right) \;,
\end{align}
where 
\begin{align}
	{\bm P} = \begin{pmatrix} p_x \\ p_y \\ p_z+e\bar{E}T \end{pmatrix} \;. 
\end{align} 
Therefore, $\Pi_{ij}$ is $T$ independent, so is $\sigma_{ij}$.

We now Fourier transform $\sigma_{ij}$ to get the electric permittivity via Eq.~(\ref{gqerqeqewq:13}).  A straightforward calculation yields
\begin{align}
	\epsilon_{ij}(\omega) 
	= 1 + \frac{1}{\omega^2} \int^{+\infty}_{-\infty}{\rm d}\tau\,{\rm e}^{+{\rm i}\omega \tau} \Theta(\tau) \Pi_{ij}(\tau) \;. \label{rqerdqdaa:34}
\end{align}
Unfortunately, it is a complicated task to fully evaluate the $\tau$ integration due to the step function, which adds one more extra frequency integration in the Fourier space and gives rise to ultraviolet divergence in general.  However, the evaluation is rather easy and feasible if we focus on the imaginary part, for which it is possible to rewrite the integrand in such a way that the step function goes away and the calculation just reduces to evaluating the Fourier transformation of $\Pi_{ij}$.  To see this, we notice
\begin{align}
	\Pi_{ij} \in {\mathbb R}\ \ {\rm and}\ \ \Pi_{ij}(\tau) = -\Pi_{ji}(-\tau) \;,
\end{align}
which immediately follow from the definition of $\Pi_{ij}$ (\ref{geqrqsa:29}).  We then find
\begin{align}
	{\rm Im}\,\epsilon_{ij}(\omega) 
	&= \frac{-{\rm i}}{\omega^2} \int^{+\infty}_{-\infty}{\rm d}\tau\,{\rm e}^{+{\rm i}\omega \tau} \nonumber\\
		&\quad \times \left[ \Pi_{(ij)}(\tau) + \left( \Theta(\tau) - \Theta(-\tau) \right) \Pi_{[ij]}(\tau) \right] \nonumber\\
	&= \frac{1}{\omega^2} {\rm Im} \, \tilde{\Pi}_{ij}(\omega) \;, \label{erqasdasda:36}
\end{align}
where $\Pi_{(ij)} := (\Pi_{ij} + \Pi_{ji})/2$ and $\Pi_{[ij]} := (\Pi_{ij} - \Pi_{ji})/2$, $\tilde{\Pi}_{ij}$ is the Fourier transformation of $\Pi_{ij}$, and we have used $\epsilon_{ij} \propto \delta_{ij}$ (\ref{qweqw:17}) in the second line to drop the anti-symmetric contribution $\Pi_{[ij]}$.

\subsection{Kramers-Kr\"{o}nig relation}  \label{sec:3C}

The real and imaginary parts of the electric permittivity are related to each other through the Kramers-Kr\"{o}nig relation, which essentially follows from causality of the electromagnetic response.  Therefore, we can immediately obtain the real part, and thereby the whole electric permittivity $\epsilon_{ij}$, once the imaginary part is computed via Eq.~(\ref{erqasdasda:36}).  Note that the real and imaginary parts are related to each other due to causality, and no other physical conditions such as unitarity are required/assumed here.

Let us show how the Kramers-Kr\"{o}nig relation arises from causality.  Causality is manifested in our in-in formulation as the step functions in the equations, e.g., $\Theta(\tau)$ in Eq.~(\ref{rqerdqdaa:34}).  To see how the step functions lead to the Kramers-Kr\"{o}nig relation, it is convenient to rewrite the integrand in Eq.~(\ref{rqerdqdaa:34}) by using the integration by parts as
\begin{align}
	\epsilon_{ij}(\omega) = 1 - \tilde{\chi}_{ij}(\omega) \;, \label{gfqerq:37}
\end{align} 
where $\tilde{\chi}$ is the Fourier transformation of
\begin{align}
	\chi_{ij}(\tau) = \int^{\tau}_{-\infty} {\rm d}\tau' \int^{\tau'}_{-\infty} {\rm d}\tau''\, \Theta(\tau'') \Pi_{ij}(\tau'') \;.  
\end{align}
The point here is that $\chi_{ij}(\tau) = 0$ for $\tau < 0$, as the integration range sets the hierarchy of the times, $\tau > \tau' > \tau''$, and the $\tau'' < 0$ contribution is vanishing due to the step function $\Theta(\tau'')$, or causality of the electric permittivity.  Therefore, 
\begin{align}
	\chi_{ij}(\tau) = \Theta(\tau) \chi_{ij}(\tau)  \label{fagqerqerq;38}
\end{align}
holds.  The Fourier transformation $\tilde{\chi}_{ij}$ must then satisfy
\begin{align}
	\tilde{\chi}_{ij}(\omega)
	&= \int^{+\infty}_{-\infty} {\rm d}\tau\,{\rm e}^{+{\rm i}\omega \tau} \Theta(\tau) \chi_{ij}(\tau) \\
	&= \int^{+\infty}_{-\infty} {\rm d}\tau\,{\rm e}^{+{\rm i}\omega \tau} \frac{1}{2\pi{\rm i}} \int^{+\infty}_{-\infty} {\rm d}\omega' \frac{{\rm e}^{+{\rm i}\omega' \tau}}{\omega' - {\rm i}0^+} \chi_{ij}(\tau) \nonumber\\
	&= \frac{1}{2\pi{\rm i}} {\rm P.V.} \int^{+\infty}_{-\infty} {\rm d}\omega' \frac{1}{\omega' - \omega} \tilde{\chi}_{ij}(\omega') + \frac{1}{2} \tilde{\chi}_{ij}(\omega) \;, \nonumber
\end{align}
where we have used the Fourier transform of the step function in the second line and the Sokhotski-Plemelj formula $1/(x - {\rm i}0^+) = {\rm P.V.}(1/x) + {\rm i}\pi \delta(x)$, with ${\rm P.V.}$ meant to take the principal value, in the last line.  Putting this relation back into Eq.~(\ref{gfqerq:37}) and taking the real part of both hand sides yield
\begin{align}
	{\rm Re}\,\epsilon_{ij}(\omega) 
	&= 1 + \frac{1}{\pi} {\rm P.V.} \int^{+\infty}_{-\infty} {\rm d}\omega' \frac{1}{\omega'-\omega} {\rm Im}\, \epsilon_{ij}(\omega') \;,  \label{gqerqq:40}
\end{align}
which is the Kramers-Kr\"{o}nig relation.  Note that the imaginary part is an odd function in $\omega$ (\ref{egqefsreqr:18}), and therefore the Kramers-Kr\"{o}nig relation (\ref{gqerqq:40}) can also be expressed as
\begin{align}
	{\rm Re}\,\epsilon_{ij}(\omega) 
	&= 1 + \frac{1}{\pi} {\rm P.V.} \int^{\infty}_{0} {\rm d}\omega' \frac{2\omega'}{\omega^{\prime 2}-\omega^2} {\rm Im}\, \epsilon_{ij}(\omega') \;,  \label{gqerqq:40--}
\end{align}
which slightly easier to be handled than Eq.~(\ref{gqerqq:40}).

We wish to apply the Kramers-Kr\"{o}nig relation (\ref{gqerqq:40--}) to obtain the real part from the imaginary part.  This, however, needs a caution.  In the above derivation, we have implicitly assumed that the imaginary part ${\rm Im}\, \epsilon_{ij}$ goes vanishing sufficiently fast (faster than $1/\omega$) at $\omega \to \infty$; otherwise the Hilbert transformation on the right-hand side of the Kramers-Kr\"{o}nig relation (\ref{gqerqq:40--}) does not exist.  Unfortunately, this is the case in our problem: the imaginary part is finite at $\omega \to \infty$ as we see shortly below [see Eq.~(\ref{eq::::8})].  Intuitively, this is because the Dirac sea is infinitely deep and therefore there always exists an electron having the excitation energy that matches the energy supplied by the probe $\omega$, once the threshold condition $\omega > 2m$ is satisfied.

To apply the Kramers-Kr\"{o}nig relation (\ref{gqerqq:40--}), we consider the difference, instead of the original value, of the electric permittivity in the presence and absence of the strong field $\Delta \epsilon_{ij}$ (\ref{qefqedqas:14}).  At $\omega \to \infty$, the energy supply by the probe field is far superior to that by the strong field.  Accordingly, the strong-field effect becomes negligible in the limit of $\omega \to \infty$, which implies $\epsilon_{ij}(\bar{\bm E}) \to \epsilon_{ij}({\bm 0})$ and hence $\Delta \epsilon_{ij} \to 0$.  Thus, ${\rm Im}\,\Delta \epsilon_{ij}$ goes vanishing sufficiently fast, to which the Kramers-Kr\"{o}nig relation (\ref{gqerqq:40--}) can safely be applied.  Note that there exist other possible ways to apply the Kramers-Kr\"{o}nig relation (\ref{gqerqq:40--}) for non-integrable functions like our electric permittivity $\epsilon_{ij}$; see Ref.~\cite{Toll:1956cya}.  We nonetheless focus on the difference in this paper, as the difference is of more interest physically and is the quantity that is usually studied in the semiconductor context.

\subsection{Evaluation of the imaginary part ${\rm Im}\,\epsilon$}  \label{sec:3D}

The imaginary part of the electric permittivity (\ref{erqasdasda:36}) can be evaluated analytically.  Simply speaking, we just have to Fourier-transform $\Pi_{ij}$.  This is doable, as $\Pi_{ij}$ can be expressed with known functions because it is made up of the mode function $\psi^{\rm in}_{\pm,{\bm p},s}$ and the mode equation (\ref{fqewdasx;30}) is analytically solvable.  The evaluation is thus straightforward but is quite involved, and therefore we put the details in Appendix~\ref{app:A}.

We have found 
\begin{align}
	&{\rm Im}\,\epsilon_\perp = {\rm Im}\,\epsilon_\parallel \nonumber\\
	&= \frac{\alpha}{3} \Theta(\omega-2m)\sqrt{1 -4\frac{m^2}{\omega^2}} \left( 1 + 2 \frac{m^2}{\omega^2} \right) \;, \label{eq::::8}
\end{align}
for $e\bar{E} = 0$ and 
\begin{subequations} \label{g2e;::15}
\begin{align}
	{\rm Im}\,\epsilon_\perp
	&= 4\pi \alpha \int_{0}^\infty \frac{{\rm d}p_\perp}{m}\frac{p_\perp}{m}  {\rm e}^{-\pi \frac{m_\perp^2}{e\bar{E}}} \Bigg[ \nonumber\\
			&\quad  +  \left( \frac{m}{\omega} \right)^2 \left( \frac{m_\perp}{\omega} \right)^2 \left| M_{\frac{1}{2} + {\rm i}\frac{m_\perp^2}{2e\bar{E}},0}\left({\rm i}\frac{\omega^2}{2e\bar{E}}\right) \right|^2  \nonumber\\
				&\quad  - \left( \frac{m}{\omega} \right)^4  {\rm Im}\left[ \left\{ M_{\frac{1}{2} + {\rm i}\frac{m_\perp^2}{2e\bar{E}},0}\left({\rm i}\frac{\omega^2}{2e\bar{E}}\right) \right\}^2 \right] \Bigg] \;, \\
	{\rm Im}\,\epsilon_\parallel
	&= 4\pi \alpha \int_{0}^\infty \frac{{\rm d}p_\perp}{m}\frac{p_\perp}{m} \nonumber\\
		&\quad \times   {\rm e}^{-\pi \frac{m_\perp^2}{e\bar{E}}}  \frac{m^2}{e\bar{E}} \left( \frac{m_\perp}{\omega} \right)^4  \left| M_{{\rm i}\frac{m_\perp^2}{2e\bar{E}},\frac{1}{2}}\left({\rm i}\frac{\omega^2}{2e\bar{E}}\right) \right|^2  \;, 
\end{align}
\end{subequations}
for $e\bar{E} \neq 0$.  The constant $\alpha := e^2/4\pi \approx 1/137$ is the fine structure constant of QED, $M_{\kappa,\mu}$ is the Whittaker function, and $m_\perp := \sqrt{m^2 + p_\perp^2}$ is the transverse mass.  The off-diagonal components $\epsilon_{ij}$ ($i \neq j$) can be shown explicitly to be vanishing, as expected from the general consideration (\ref{qweqw:17}).  Note that we have assumed $\omega>0$.  The values for $\omega < 0$ is obtained by simply multiplying $(-1)$; see Eq.~(\ref{egqefsreqr:18}).

We must discuss the infrared (i.e., low-frequency $\omega \to 0$) behavior of $\epsilon_{ij}$ (\ref{g2e;::15}).  Taking $\omega \to 0$ yields
\begin{subequations} \label{eq:::::::a8}
\begin{align}
	&{\rm Im}\,\epsilon_\perp \nonumber\\
	&\to \frac{\alpha}{\pi} {\rm e}^{-\pi \frac{m^2}{e\bar{E}}} \left[ \frac{e\bar{E}}{\omega^2} +  \left( \frac{1}{\pi} + \frac{1}{2} \frac{m^2}{e\bar{E}} \right) + {\mathcal O}(\omega^2) \right] \;, \\
	&{\rm Im}\,\epsilon_\parallel \nonumber\\
	&\to \frac{\alpha}{\pi} {\rm e}^{-\pi \frac{m^2}{e\bar{E}}} \left[  \left( \frac{1}{\pi} + \frac{m^2}{e\bar{E}} + \frac{\pi}{2} \left(\frac{m^2}{e\bar{E}}\right)^2 \right) + {\mathcal O}(\omega^2) \right]
\;. 
\end{align}
\end{subequations}
Thus, the (transverse) electric permittivity is divergent at $\omega \to 0$.  Remark that the divergence is proportional to the non-perturbative factor ${\rm e}^{-\pi m^2/e\bar{E}}$, meaning that the infrared divergence is a strong-field effect.  Such a non-perturbative effect is dismissed in the weak-field limit [in fact, the free-field result $\epsilon(\omega;\bar{\bm E}={\bm 0})$ (\ref{eq::::8}) is finite at $\omega \to 0$].

The infrared divergence implies that multi-photon processes of ${\mathcal O}(|{\bm {\mathcal E}}|^n)$ ($n>1$) need to be resummed.  Indeed, when the Keldysh parameter for the probe field $\gamma = m\omega/|e{\bm {\mathcal E}}|$, which controls the non-linearity of QED, becomes large, the lowest-order treatment is known to be invalidated~\cite{1970PhRvD...2.1191B, Popov:1971iga, Taya:2014taa, Taya:2020dco, Huang:2019uhf}.  The resummation would subtract the infrared divergence, and thereby leads to a physically-meaningful number for the linear electric permittivity at the low-frequency regime, as is the case in the usual quantum-field theoretic calculations of, e.g., cross section.  The subtraction is also important for our Kramers-Kr\"onig approach, as the Hilbert transformation does not converge for any $\omega$ if the integrand (\ref{gqerqq:40--}) contains singularities on the integration axis.  In principle, what we need to do then is to resum higher-order processes and to see the cancellation among them how to yield an infrared-safe linear electric permittivity.  We, unfortunately, found it difficult, as the infrared behavior is significantly affected by the non-perturbative strong-field effects and the standard resummation procedure established in the free-field limit (e.g., Refs.~\cite{Peskin:1995ev, Weinberg:1995mt}) cannot be applied.  (Note that although the removal of the infrared divergence is theoretically important and is essential to predict physical values at the very low-frequency regime $\omega \approx 0$ as explained above, it is also true that it can give only negligible effects for most values of $\omega \gtrsim 0$ and thus the bare formula (\ref{g2e;::15}) is quite sufficient in practice; see the later discussions around Fig.~\ref{fig1}.  )

We, therefore, deal with the infrared divergence rather phenomenologically, without explicitly carrying out the difficult resummation (which we leave as future work).  We shall explain the phenomenological idea to remove the infrared divergence in the next two subsections.  Briefly, we remove the infrared divergence by introducing a counter term, which is added by hand though presumably appears as a result of the resummation.  We fix the value of the counter term appropriately by making use of that the imaginary part of the electric permittivity and the number of pairs produced by the dynamically-assisted Schwinger effect are in exact agreement with each other and by regularizing the number of pairs using the established formula for the Schwinger effect in the low-frequency limit.

\subsection{Relation to the dynamically assisted Schwinger effect}  \label{sec:3E}

We show that the imaginary part of the electric permittivity of the vacuum exactly agrees with the number of pairs produced by the dynamically-assisted Schwinger effect (i.e., the electron-positron pair production from the vacuum in a constant electric field superimposed by a time-dependent field), up to some unimportant factors.  

To establish the relation between the number of pairs and the imaginary part of the electric permittivity, we begin with reminding the phenomenological argument of electromagnetism in matter that a non-zero imaginary part ${\rm Im}\,\epsilon_{ij} \neq 0$ means that the injected probe ${\bm {\mathcal E}}$ dissipates energy, which is the so-called dielectric loss.  The rate of the dielectric energy loss per unit spacetime volume $W$ for the probe with the monochromatic frequency $\omega$ (\ref{eq:2}) is given by (see, e.g., Ref.~\cite{lan84, STERN1963299})
\begin{align}
	W = \frac{1}{2} \omega {\mathcal E}_{i} {\mathcal E}_{j} \,{\rm Im}\,\epsilon_{ij} \;. \label{eq;27}
\end{align}
The dielectric energy loss (\ref{eq;27}) is, from a microscopic viewpoint, caused by the decay of the probe field ${\bm {\mathcal E}}$ due to particle production.  As energy $\omega$ is dissipated whenever the probe field decays, we have
\begin{align}
	W = \omega \Gamma \;, \label{eq;28}
\end{align}
where $\Gamma$ is the decay rate of the probe field per unit spacetime volume.  Therefore, we obtain
\begin{align}
	\Gamma = \frac{1}{2} {\mathcal E}_{i} {\mathcal E}_{j} \,{\rm Im}\,\epsilon_{ij} \;. \label{fqerqwdq:47}
\end{align}

We conjecture that the same relation (\ref{fqerqwdq:47}) holds in QED.  As a pair is produced whenever the probe decays, the decay rate $\Gamma$ may be read off from the number of produced pairs via the dynamically-assisted Schwinger effect $N$:
\begin{align}
    \Gamma = \frac{N({\bm {\mathcal E}} \neq {\bm 0}) - N({\bm {\mathcal E}} = {\bm 0})}{V_4} \;, \label{eq:24}
\end{align}
where $V_4 = \int {\rm d}^4x$ is the whole spacetime volume of the system.  We have subtracted the ${\bm {\mathcal E}} = {\bm 0}$ contribution to remove the Schwinger contribution without the probe, since it is irrelevant to the decay of the probe.  Expanding Eq.~(\ref{eq:24}) in ${\bm {\mathcal E}}$ and then plugging it back into Eq.~(\ref{fqerqwdq:47}), we arrive at
\begin{align}
	{\rm Im}\,\epsilon_{ij} = \frac{1}{V_4} \left. \frac{\partial^2 N}{\partial {\mathcal E}_{i} \partial {\mathcal E}_{j} } \right|_{{\bm {\mathcal E}} = {\bm 0}} \;. \label{eq;vwefewqf50}
\end{align}
This is the relation connecting the dynamically-assisted Schwinger effect and the imaginary part of the electric permittivity.

We remark that Eq.~(\ref{eq;vwefewqf50}) shows that the imaginary part is {\it not} directly related to the scattering amplitude (or its squared) of the in-out formalism.  In fact, the number of pairs $N$, which is, in a quantum-field theoretical language, defined as an in-in vacuum expectation value of the electron number operator at the asymptotic out-state $a^{{\rm out}\dagger}_{{\bm p},s} a^{\rm out}_{{\bm p},s}$\footnote{The number of pairs and that of electrons must be the same because of the gauge invariance.  Therefore, we do not carefully distinguish the two quantities in this work.
}:
\begin{align}
	N := \sum_s \int {\rm d}^3{\bm p} \braket{{\rm vac;in}| a^{{\rm out}\dagger}_{{\bm p},s} a^{\rm out}_{{\bm p},s} |{\rm vac;in}} \;. \label{gefdafafa:51}
\end{align}
This quantity should be clearly distinguished from the squared amplitudes (i.e., the probabilities) to produce pairs such as $|\braket{ e^+ e^-;{\rm out}| {\rm vac;in}}|^2$ in the in-out formalism.  In general, an in-in expectation value is identical to the sum of in-out amplitudes.  For the number of pairs $N$ (\ref{gefdafafa:51}), inserting the complete set $1 = \sum_{X \in {\rm all\; states}} \ket{ X;{\rm out} } \bra{ X;{\rm out} } $ inside of the number operator, we find
\begin{align}
	N = \sum_{X \in {\rm all\; states}} \sum_s \int {\rm d}^3{\bm p} \left| \braket{ Xe^{-}_{{\bm p},s}; {\rm out} | {\rm vac;in} } \right|^2 \;. \label{gqerqer:52}
\end{align}
Using the Bogoliubov-transformation technique, it can be shown that the vacuum at the in-state equals to the superposition of all the pair states at the out-state (see Ref.~\cite{Tanji:2008ku} for details): 
\begin{align}
	\ket{ {\rm vac;in} }
	&= \prod_{{\bm p},s} \left[ 1 + ({\rm const}) \times a^{{\rm out}\dagger}_{{\bm p},s} b^{{\rm out}\dagger}_{-{\bm p},-s} \right] \ket{{\rm vac;out}} \;,
\end{align}
where $b^{\rm out}_{{\bm p},s}$ is the out-state positron annihilation operator.  Therefore, the non-vanishing contributions to Eq.~(\ref{gqerqer:52}) are 
\begin{align}
	X =\;	&e^{+}_{-{\bm p},-s} \;, \nonumber\\
			&e^{+}_{-{\bm p},-s} e^{+}_{-{\bm p}',-s'}e^{-}_{{\bm p}',s'} \;, \nonumber\\
			&e^{+}_{-{\bm p},-s} e^{+}_{-{\bm p}',-s'}e^{-}_{{\bm p}',s'} e^{+}_{-{\bm p}'',-s''}e^{-}_{{\bm p}'',s''} \;, \cdots \;.
\end{align}
Substituting this into Eq.~(\ref{gefdafafa:51}) yields
\begin{align}
	N = \sum_{X' \in {\rm all\; pairs}} \sum_s \int {\rm d}^3{\bm p} \left| \braket{ X'; {\rm out} | {\rm vac;in} } \right|^2 \;.  \label{gqerqereq:55}
\end{align}
Thus, $N$ corresponds to the sum of {\it all} the possible squared amplitudes to produce pairs.  In the weak-field limit, the multiple pair-production events may be suppressed and thus the number $N$ can be approximately equal to the squared amplitude for the single pair production.  In general, however, this is not necessarily the case, e.g., for supercritical fields $e\bar{E} \gtrsim m^2$.

At the moment, the relation (\ref{eq;vwefewqf50}) is just a conjecture.  We now turn to prove Eq.~(\ref{eq;vwefewqf50}) by explicitly calculating the number of pairs (\ref{gefdafafa:51}) using the Furry-picture approach for the dynamically-assisted Schwinger effect (see Refs.~\cite{Torgrimsson:2017pzs, Taya:2018eng, Torgrimsson:2018xdf, Huang:2019uhf, Taya:2020pkm, Taya:2020bcd} for details).  Similarly to the in-state annihilation operators (\ref{fqewqw:21}), the out-state ones, $a_{{\bm p},s}^{\rm out}$ and $b_{{\bm p},s}^{\rm out}$, are defined by expanding the field operator $\psi^{\rm out} := \lim_{t \to \infty} \psi$ with the appropriate out-state mode function $\psi^{{\rm out}}_{\pm,{\bm p},s}$.  The out-state mode function $\psi^{{\rm out}}_{\pm,{\bm p},s}$ is identified, similarly to $\psi^{{\rm in}}_{\pm,{\bm p},s}$, as a solution to the same mode equation (\ref{fqewdasx;30}) but with a different boundary condition set at $t \to +\infty$: 
\begin{align}
	&\lim_{t \to +\infty} \begin{pmatrix} \psi^{\rm out}_{+,{\bm p},s} \\ \psi^{\rm out}_{-,{\bm p},s} \end{pmatrix} \nonumber\\
	&= \begin{pmatrix} u_{{\bm p}-e\bar{\bm A}(+\infty),s} \\ v_{-{\bm p}-e\bar{\bm A}(+\infty),s} \end{pmatrix} {\rm e}^{\mp {\rm i}\sqrt{m^2 + ({\bm p}-e\bar{\bm A}(+\infty))^2}t} \;.   \label{aaaeq-10}
\end{align}
Then, using the orthonormality of the mode function $\psi^{{\rm out}\dagger}_{\pm,{\bm p},s} \psi^{{\rm out}}_{\pm',{\bm p},s'} = \delta_{\pm,\pm'} \delta_{s,s'}$ and the perturbative solution to the Dirac equation (\ref{fqewrqe:435}), we find
\begin{align}
	&\begin{pmatrix} a^{\rm out}_{{\bm p},s} \\ b^{{\rm out}\dagger}_{-{\bm p},-s} \end{pmatrix}
	= \int {\rm d}^3{\bm x} \frac{{\rm e}^{-{\rm i}{\bm p}\cdot{\bm x}}}{(2\pi)^{3/2}} \psi^{{\rm out}\dagger}_{\pm, {\bm p},s}(t) \psi^{\rm out}(t,{\bm x}) \\
	&= \lim_{t \to \infty} \int {\rm d}^3{\bm x} \frac{{\rm e}^{-{\rm i}{\bm p}\cdot{\bm x}}}{(2\pi)^{3/2}} \psi^{{\rm out}\dagger}_{\pm, {\bm p},s}(t) \Big[ \psi^{\rm in}(t,{\bm x}) \nonumber\\
		&\quad + e\int{\rm d}^4x'\, S^{\rm R}(t,t'; {\bm x},{\bm x}') \slashed{\mathcal A}(t') \psi^{\rm in}(t',{\bm x}')   +   {\mathcal O}(|{\bm {\mathcal E}}|^2) \Big] \;. \nonumber
\end{align}
Inserting this into the number of pairs (\ref{gefdafafa:51}), we obtain, at the leading order in the probe ${\bm {\mathcal E}}$, that
\begin{align}
	N
	&= V_3 \sum_{s,s'} \int \frac{{\rm d}^3{\bm p}}{(2\pi)^3} \label{eq:13} \\ 
		&\quad \times \left| \psi^{{\rm out}\dagger}_{+,{\bm p},s} \psi^{\rm in}_{-,{\bm p},s'} - {\rm i}e \int {\rm d}t \, \bar{\psi}^{{\rm out}}_{+,{\bm p},s} \slashed{\mathcal A} \psi^{{\rm in}}_{-,{\bm p},s'} \right|^2 \;.  \nonumber
\end{align}
The calculation of the number of pairs is thus reduced to evaluating the overlap between in- and out-state mode functions.  As the mode equation (\ref{fqewdasx;30}) is solvable, $\psi^{{\rm in}}_{\pm,{\bm p},s}$ and $\psi^{{\rm out}}_{\pm,{\bm p},s}$ can be expressed with known functions and hence it is possible to analytically evaluate Eq.~(\ref{eq:13}) (see Appendix~\ref{app:B} for more details).  We have found 
\begin{align}
	\frac{N}{V_4}
	&= 2\pi \alpha \int_{0}^\infty \frac{{\rm d}p_\perp}{m}\frac{p_\perp}{m}  {\rm e}^{-\pi \frac{m_\perp^2}{e\bar{E}}} \Bigg[ \frac{e\bar{E}}{m^2} \nonumber\\
			&\quad  +  \left( \frac{m}{\omega} \right)^2 \left( \frac{m_\perp}{\omega} \right)^2 \left| M_{\frac{1}{2} + {\rm i}\frac{m_\perp^2}{2e\bar{E}},0}\left({\rm i}\frac{\omega^2}{2e\bar{E}}\right) \right|^2  {\mathcal E}_\perp^2 \nonumber\\
				&\quad  - \left( \frac{m}{\omega} \right)^4  {\rm Im}\left[ \left\{ M_{\frac{1}{2} + {\rm i}\frac{m_\perp^2}{2e\bar{E}},0}\left({\rm i}\frac{\omega^2}{2e\bar{E}}\right) \right\}^2 \right]  {\mathcal E}_\perp^2 \nonumber\\
				&\quad + \frac{m^2}{e\bar{E}} \left( \frac{m_\perp}{\omega} \right)^4  \left| M_{{\rm i}\frac{m_\perp^2}{2e\bar{E}},\frac{1}{2}}\left({\rm i}\frac{\omega^2}{2e\bar{E}}\right) \right|^2  {\mathcal E}_z^2  \Bigg] \;. \label{gqerqewqrwq:60}
\end{align}
where ${\mathcal E}_\perp^2 := \sqrt{{\mathcal E}_x^2+{\mathcal E}_y^2}$.  Comparing this with the imaginary part (\ref{g2e;::15}), we arrive at
\begin{align}
	\frac{N(\bar{E})}{V_4} 
	&= \frac{N(\bar{E}=0)}{V_4}  + \frac{1}{2} {\mathcal E}_\perp^2 \, {\rm Im}\,\epsilon_{\perp} + \frac{1}{2} {\mathcal E}_z^2 \, {\rm Im}\,\epsilon_{\parallel} \;,
\end{align}
which proves the relation (\ref{eq;vwefewqf50}).

\subsection{Removal of the infrared divergence}  \label{sec:3F}

We utilize the exact correspondence between the dynamically-assisted Schwinger effect and the imaginary part of the electric permittivity (\ref{eq;vwefewqf50}) to remove the infrared divergence of the electric permittivity.  The correspondence implies that the number of pairs $N$ also suffers from the infrared divergence, and it should be regularized consistently with the imaginary part.  It is easier to regularize the number of pairs $N$, instead of directly dealing with the imaginary part, as there exists an established treatment of the Schwinger effect for slowly varying fields.  Namely, the locally-constant-field approximation~\cite{Bulanov:2004de, Dunne:2005sx, Dunne:2006st, Aleksandrov:2018zso}, by which the number of pairs in the low-frequency limit can be computed in a well-defined manner without divergence.

To be concrete, we phenomenologically introduce a counter term $\Delta N$, which presumably comes as a result of the resummation, to the bare number (\ref{gqerqewqrwq:60}) as
\begin{align}
	N^{\rm reg} := N - \Delta N \;. \label{fcqf;10}
\end{align}
The counter term $\Delta N$ is determined by matching the physical number $N^{\rm reg}$ to the locally-constant-field-approximation result [Eq.~(\ref{eq::12}) below], which gives the accurate number in the low-frequency limit.  Once the number of pairs regularized, we can naturally obtain the regularized imaginary part using the correspondence (\ref{eq;vwefewqf50}) as
\begin{align}
	{\rm Im}\,\epsilon_{ij}^{\rm reg} 
	= \frac{1}{V_4} \left. \frac{\partial^2 N^{\rm reg}}{\partial {\mathcal E}_{i} \partial {\mathcal E}_{j} } \right|_{{\bm {\mathcal E}} = {\bm 0}} \;. \label{eq;vwefewqf50--}
\end{align}

Let us illustrate the matching procedure in detail.  We first remind that the number of produced particles by a constant electric field ${\bm E}$ is given by the Schwinger(-Nikishov) formula~\cite{Schwinger:1951nm, Nikishov:1969tt}: 
\begin{align}
	N^{\rm Schwinger} = V_4 \frac{|e{\bm E}|^2}{4\pi^3} {\rm e}^{-\pi \frac{m^2}{|e{\bm E}|}} \;.
\end{align}
The Schwinger formula is exact for a constant electric field and may well be applied to slow electric field ${\bm E} \to {\bm E}(t)$ with frequency $\omega \to 0$ (the locally-constant-field approximation).  Namely, so long as the typical time scale for the variation of the field $\sim 1/\omega$ is much slower compared to that for a pair production to take place $\sim m/|e{\bm E}|$, a pair production may be regarded as an instantaneous process that occurs locally at each instant of time $t$ with a constant electric field with the instantaneous value ${\bm E}(t)$.  Thus, 
\begin{align}
	N^{\rm LCFA} = V_3 \int {\rm d}t \frac{|e{\bm E}(t)|^2}{4\pi^3} {\rm e}^{-\pi \frac{m^2}{|e{\bm E}(t)|}} \;, \label{eq::12}
\end{align}
with $V_3 := \int {\rm d}^3{\bm x}$ being the spatial volume, gives the correct number of pairs in the low-frequency limit $\omega \to 0$.  For our field configuration (\ref{eq:2}), the locally-constant-field approximation (\ref{eq::12}) yields, after expanding with respect to the probe ${\bm {\mathcal E}}$, that
\begin{align}
	&N^{\rm LCFA} \label{eq::::a8} \\
	&= V_4 \frac{(e\bar{E})^2}{4\pi^3} {\rm e}^{-\pi \frac{m^2}{e\bar{E}}} \Bigg[ 1 +  \left( \frac{1}{2} + \frac{\pi}{4} \frac{m^2}{e\bar{E}} \right)  \left(  \frac{ {\mathcal E}_\perp }{\bar{E}} \right)^2   \nonumber\\
		&\quad  +  \left( \frac{1}{2} +  \frac{\pi}{2} \frac{m^2}{e\bar{E}} + \frac{\pi^2}{4} \left( \frac{m^2}{e\bar{E}} \right)^2 \right)  \left( \frac{ {\mathcal E}_z }{\bar{E}}  \right)^2   +  {\mathcal O}(\omega^2, |{\bm {\mathcal E}}|^4) \Bigg] \;. \nonumber
\end{align}

Our idea to fix the counter term $\Delta N$ is to require that the regularized number $N^{\rm reg}$ (\ref{fcqf;10}) reproduces the locally-constant-field-approximation result $N^{\rm LCFA}$ (\ref{eq::::a8}) at $\omega \to 0$.  Then, noticing
\begin{align}
	N \xrightarrow{\omega \to 0} \; 
	& V_4 \frac{(e\bar{E})^2}{4\pi^3} {\rm e}^{-\pi \frac{m^2}{e\bar{E}}} \Bigg[ 1  +  \frac{\pi}{2}  \frac{e\bar{E}}{\omega^2} \left(  \frac{ \sqrt{ {\mathcal E}_x^2 + {\mathcal E}_y^2} }{\bar{E}} \right)^2  \label{eq:::::aaa::a8} \\
		&  +  \left\{  \left( \frac{1}{2} + \frac{\pi}{4} \frac{m^2}{e\bar{E}} \right)  \left(  \frac{ \sqrt{ {\mathcal E}_x^2 + {\mathcal E}_y^2} }{\bar{E}} \right)^2  \right. \nonumber\\
		& \left. +  \left( \frac{1}{2} +  \frac{\pi}{2} \frac{m^2}{e\bar{E}} + \frac{\pi^2}{4} \frac{m^4}{(e\bar{E})^2} \right)  \left( \frac{ {\mathcal E}_z }{\bar{E}}  \right)^2  \right\}  +  {\mathcal O}(\omega^2) \Bigg] \;, \nonumber
\end{align}
we can uniquely fix $\Delta N$ as
\begin{align}
    \Delta N 
    &= V_4 \frac{(e\bar{E})^2}{4\pi^3} {\rm e}^{-\pi \frac{m^2}{e\bar{E}}} \frac{\pi}{2}  \frac{e\bar{E}}{\omega^2} \left(  \frac{{\mathcal E}_{\perp} }{\bar{E}} \right)^2 \; . \label{eq::14} 
\end{align}

Therefore, from the correspondence (\ref{eq;vwefewqf50--}), the imaginary part ${\rm Im}\,\epsilon^{\rm reg}_{ij}$ should be regularized as 
\begin{subequations} \label{g2e;::15ffff}
\begin{align}
	{\rm Im}\,\epsilon^{\rm reg}_\perp
	&= {\rm Im}\,\epsilon_\perp - \frac{\alpha}{\pi} {\rm e}^{-\pi \frac{m^2}{e\bar{E}}} \frac{e\bar{E}}{\omega^2} \;, \\
	{\rm Im}\,\epsilon^{\rm reg}_\parallel
	&= {\rm Im}\,\epsilon_\parallel  \;.  
\end{align}
\end{subequations}
That is, we just need to subtract the most divergent part $\propto \omega^{-2}$.  It is evident that the high-frequency behavior is unmodified by the subtraction.

\begin{figure}[!t]
\includegraphics[clip, width=0.49\textwidth]{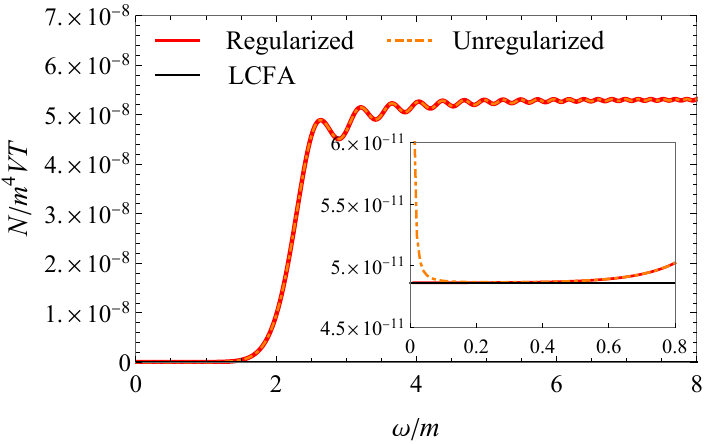}
\caption{\label{fig1} The number of pairs produced $N$ as a function of the probe frequency $\omega$.  The red line corresponds to the regularized result $N^{\rm reg}$ (\ref{fcqf;10}), which is compared with the unregularized result $N$ (\ref{gqerqewqrwq:60}) (dotted-dashed orange) and the locally-constant-field approximation $N^{\rm LCFA}$ (\ref{eq::::a8}) (solid black).  The parameters are $e\bar{E}/m^2 = 0.2$, ${\mathcal E}_\parallel/\bar{E} = 0$, and ${\mathcal E}_\perp/\bar{E} = 0.01$.  
}
\end{figure}

Before proceeding, let us mention that the infrared divergence and the regularization procedure can affect only the very low-frequency regime $\omega \approx 0$.  This is illustrated in Fig.~\ref{fig1}, in which the bare and regularized numbers, (\ref{gqerqewqrwq:60}) and (\ref{fcqf;10}), respectively, are plotted against $\omega$.  It implies that {\it in practice} it is sufficient to use the bare $N$ (\ref{gqerqewqrwq:60}), or Eq.~(\ref{g2e;::15}) for the imaginary part, when making a comparison with experiments.  In fact, experimentally, it is not always easy to realize very low-frequency fields, since $\omega = 0$ is just an idealistic limit of theory.  Let us make a quantitative estimate of when the infrared divergence can be significant.  By taking the ratio between the counter term $\Delta N$ (\ref{eq::14}) and the locally-constant-field-approximation result $N^{\rm LCFA}$ (\ref{eq::::a8}), 
\begin{align}
    \frac{\Delta N}{N^{\rm LCFA}} = \frac{\pi}{2}  \frac{e\bar{E}}{\omega^2} \left( \frac{ {\mathcal E}_\perp }{\bar{E}} \right)^2 \;,
\end{align}
we understand that the infrared divergence can be manifest when
\begin{align}
    \omega \lesssim \sqrt{\frac{\pi}{2}} \frac{ e{\mathcal E}_\perp }{\sqrt{ e\bar{E}}} \;. \label{eqvqe16}
\end{align}
This is formally vanishing in the limit of weak probes $|{\bm {\mathcal E}}| \to {\bm 0}$.  In other words, in order for the infrared divergence to be significant for finite $\omega$, the probe ${\bm {\mathcal E}}$ needs to be rather strong (at least comparable to the strong field $\bar{E}$).

\subsection{Final formula for the change of the electric permittivity}  \label{sec:3G}

Let us summarize our formula for the change of the electric permittivity $\Delta \epsilon_{ij}$ under the constant strong electric field (\ref{eq:2}).  

The imaginary part after removing the infrared divergence is, from Eqs.~(\ref{g2e;::15ffff}) and (\ref{eq::::8}), given by
\begin{subequations} \label{g2e;::15ffefq}
\begin{align}
	{\rm Im}\,\epsilon^{\rm reg}_\perp
	&= 4\pi \alpha \int_{0}^\infty \frac{{\rm d}p_\perp}{m}\frac{p_\perp}{m}  {\rm e}^{-\pi \frac{m_\perp^2}{e\bar{E}}} \Bigg[ \nonumber\\
			&\quad  +  \left( \frac{m}{\omega} \right)^2 \left( \frac{m_\perp}{\omega} \right)^2 \left| M_{\frac{1}{2} + {\rm i}\frac{m_\perp^2}{2e\bar{E}},0}\left({\rm i}\frac{\omega^2}{2e\bar{E}}\right) \right|^2  \nonumber\\
				&\quad  - \left( \frac{m}{\omega} \right)^4  {\rm Im}\left[ \left\{ M_{\frac{1}{2} + {\rm i}\frac{m_\perp^2}{2e\bar{E}},0}\left({\rm i}\frac{\omega^2}{2e\bar{E}}\right) \right\}^2 \right] \Bigg] \nonumber\\
				&\quad - \frac{\alpha}{\pi} {\rm e}^{-\pi \frac{m^2}{e\bar{E}}} \frac{e\bar{E}}{\omega^2} \nonumber\\
				&\quad - \frac{\alpha}{3} \Theta(\omega-2m)\sqrt{1 -4\frac{m^2}{\omega^2}} \left( 1 + 2 \frac{m^2}{\omega^2} \right) \;, \\
	{\rm Im}\,\epsilon^{\rm reg}_\parallel
	&= 4\pi \alpha \int_{0}^\infty \frac{{\rm d}p_\perp}{m}\frac{p_\perp}{m} \nonumber\\
		&\quad \times   {\rm e}^{-\pi \frac{m_\perp^2}{e\bar{E}}}  \frac{m^2}{e\bar{E}} \left( \frac{m_\perp}{\omega} \right)^4  \left| M_{{\rm i}\frac{m_\perp^2}{2e\bar{E}},\frac{1}{2}}\left({\rm i}\frac{\omega^2}{2e\bar{E}}\right) \right|^2 \nonumber\\
		&\quad - \frac{\alpha}{3} \Theta(\omega-2m)\sqrt{1 -4\frac{m^2}{\omega^2}} \left( 1 + 2 \frac{m^2}{\omega^2} \right) \;. 
\end{align}
\end{subequations}
The transverse-momentum $p_\perp$ integration can be done numerically.  Note that in the semiconductor case, the imaginary part is given in terms of the Airy function, rather than the Whittaker function~\cite{PhysRev.147.554, PhysRev.153.972}.  This is essentially because QED is relativistic, while it is non-relativistic in semiconductor, and therefore they have different dispersions.  Nevertheless, from a mathematical viewpoint, those two functions are similar to each other in the sense that both functions obey differential equations with a single turning point.  It is a general feature of such differential equations that the behavior of the solution changes from/to oscillation to/from damping at the turning point.  This is the mathematical origin of the characteristic oscillating pattern in the electric permittivity (or the Franz-Keldysh effect), which we shall discuss in detail in the next section.

The real part can be obtained from the imaginary part (\ref{g2e;::15ffefq}) by using the Kramers-Kr\"{o}nig relation: 
\begin{align}
	{\rm Re}\,\Delta\epsilon^{\rm reg}_{ij}(\omega) 
	&= \frac{1}{\pi} {\rm P.V.} \int^{\infty}_{0} {\rm d}\omega' \frac{2\omega'}{\omega^{\prime 2}-\omega^2} {\rm Im}\, \Delta\epsilon^{\rm reg}_{ij}(\omega') \;.  \label{gqerqq:40-aa-}
\end{align}
The frequency $\omega'$ integration can be done numerically.  Note that the Kramers-Kr\"{o}nig relation holds equally even after regularizing the infrared divergence.  Indeed, the subtraction of the divergent $\omega^{-2}$ term is equivalent to shifting $\tilde{\Pi}(\omega)$ in Eq.~(\ref{erqasdasda:36}) by a constant.  In the coordinate space, this means that $\Pi(\tau)$ is shifted by a delta function $\delta(\tau)$.  The proof of the Kramers-Kr\"{o}nig relation in Sec.~\ref{sec:3C} remains valid, since the essence of the proof is the step function dependence of $\chi(\tau)$ (\ref{fagqerqerq;38}), which is unmodified by the shift of $\Pi(\tau)$.

\section{Result} \label{sec:4}

We turn to discuss the physics of the electric permittivity in a strong constant electric field based on the theoretical foundation in Sec.~\ref{sec:3}.  We discuss the imaginary part ${\rm Im}\,\epsilon_{ij}$ in Sec.~\ref{sec:4a} and then the real part ${\rm Re}\,\epsilon_{ij}$ in Sec.~\ref{sec:4b} by numerically carrying out the integrations in Eqs.~(\ref{g2e;::15ffefq}) and (\ref{gqerqq:40-aa-}).  

For notational simplicity, we do not explicitly write the superscript ``reg'' in what follows.  All the numerical results presented here are the regularized values.  

\subsection{Imaginary part ${\rm Im}\,\Delta \epsilon_{ij}$} \label{sec:4a}

\begin{figure*}[!t]
\includegraphics[clip, width=0.495\textwidth]{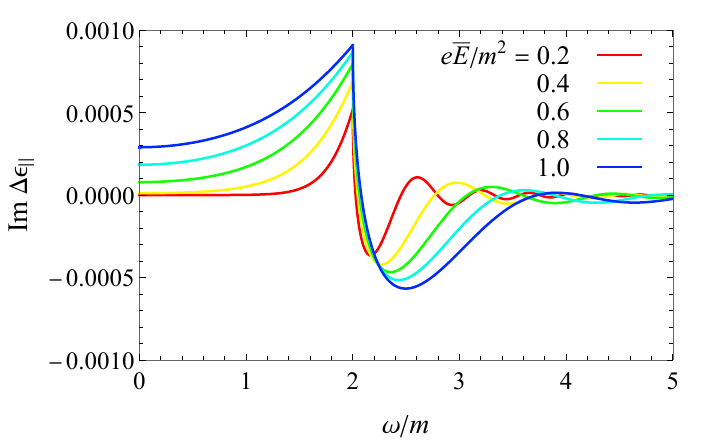}
\includegraphics[clip, width=0.495\textwidth]{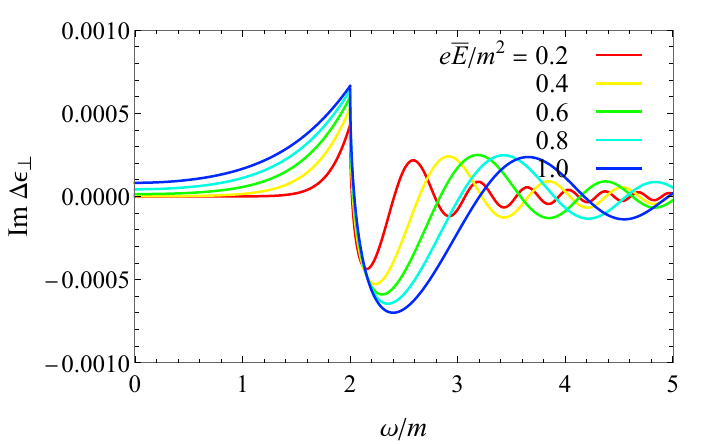}
\caption{\label{fig2} The change of the imaginary part of the longitudinal and transverse electric permittivities, ${\rm Im}\,\Delta\epsilon_\parallel$ (left) and ${\rm Im}\,\Delta\epsilon_\perp$ (right), as functions of the probe frequency $\omega$.  The color distinguishes the strength of the strong field $e\bar{E}/m^2$.  Note that if $m$ is taken to be the usual electron mass $m = 511\;{\rm keV}$ in the natural units, in the physical units it corresponds to $m = 1.24 \times 10^{20}\;{\rm Hz}$ and $m^2 = 1.32 \times 10^{18}\;{\rm V/m}$.  }
\end{figure*}

Figure~\ref{fig2} shows the frequency $\omega$ dependence of the imaginary part ${\rm Im}\,\Delta\epsilon_{ij}$ (\ref{g2e;::15ffefq}).  It exhibits two characteristic features (i) the exponential tail below the band-gap energy $\omega < 2m$ and (ii) the oscillating behavior above $\omega > 2m$.  Those are the QED analog of the electroreflectance in semiconductor, where very similar features, including the order of the sign and magnitude of the oscillation, have been observed, e.g., with Si~\cite{PhysRev.142.445}.

Those characteristic features are directly related to the change of the QED-vacuum (or the Dirac-sea) structure by the strong electric field~\cite{Taya:2018eng}.  Namely, the strong electric field $\bar{\bm E}$ tilts the QED vacuum in the direction of the strong field ${\bf e}_z$.  As a consequence, the electrons in the Dirac sea can tunnel into the gap, inside of which less energy is needed for the probe to excite a pair and thus the pair production is enhanced.  More particle production means more dielectric energy loss, and hence the electric permittivity acquires larger imaginary part compared to the free-field limit, which is the origin of the exponential tail below the band-gap energy (i).  On the other hand, that the quantum tunneling into the gap occurs means that the complementary process of the quantum tunneling, i.e., quantum refection by the gap, must occur.  Accordingly, the electrons in the Dirac sea interfere with those reflected by the gap.  This quantum interference leads to an oscillating distribution of the Dirac-sea electrons.  The pair production is likely to occur from where the Dirac-sea electrons exist more, and the excitation energy $\omega$ that the probe needs to supply for pair production is different, depending on where an electron is located in the Dirac sea.  Thus, the resulting number of pairs produced, and therefore the imaginary part, shows an oscillating dependence in $\omega$ (ii).  Note that an oscillatory dependence similar to ours (similar but has different features such as square-root divergences at thresholds) has also been obtained for strong magnetic fields~\cite{Hattori:2012je, Hattori:2020htm}, which is also induced by the change of the QED vacuum but is due to the Landau quantization of the electron energy level.

As the change of the QED vacuum by the strong field $\bar{\bm E}$ is the essence of the two features, the direction of the probe is not crucial here and thus ${\rm Im}\,\Delta \epsilon_\parallel$ and ${\rm Im}\,\Delta \epsilon_\perp$ basically have the same $\omega$ dependencies.  Nonetheless, a closer comparison shows that they slightly deviate from each other, which is the manifestation of the vacuum birefringence by the strong electric field.

\begin{figure*}[!t]
\hspace*{8.2mm}\includegraphics[clip, width=0.445\textwidth]{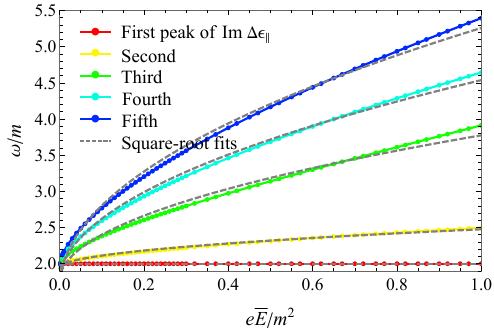}
\hspace*{8.8mm}\includegraphics[clip, width=0.445\textwidth]{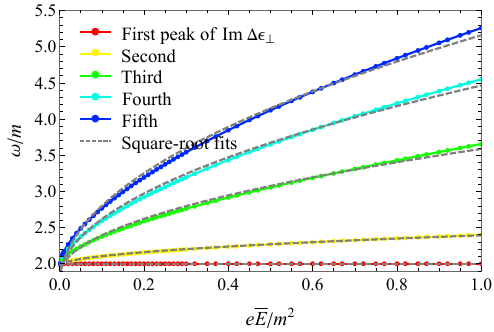} \\
\hspace*{2mm}
\includegraphics[clip, width=0.48\textwidth]{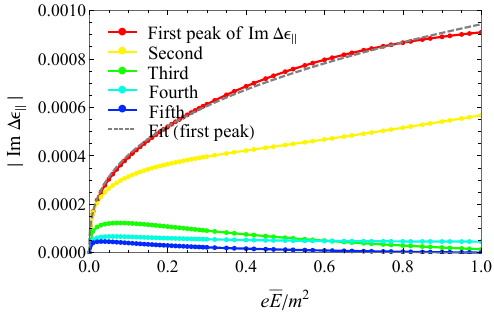}
\hspace*{2mm}
\includegraphics[clip, width=0.48\textwidth]{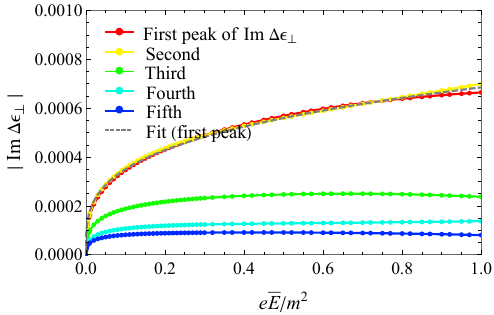}
\caption{\label{fig3} The peak locations (with respect to $\omega$) (top) and heights (bottom) of the change of the electric permittivities, ${\rm Im}\,\Delta\epsilon_\parallel$ (left) and ${\rm Im}\,\Delta\epsilon_\perp$ (right), plotted against the strength of the strong field $e\bar{E}/m^2$.  The gray dashed lines show fitting results of the curves: a square-root fit $\omega/m = c_1 + c_2 \sqrt{e\bar{E}/m^2}$ (\ref{gqewdqwedwq:73}) for each peak location on the top and a power-function fit ${\rm Im}\,\Delta \epsilon_{ij} = d_1 (e\bar{E}/m^2)^{d_2}$ (\ref{gqerqe:74}) for the heights of the largest first peaks on the bottom.  The best-fit parameters $(c_1, c_2)$ were found to be $(2.00, 0.00), (1.94, 0.536), (1.75, 2.03), (1.73, 2.82), (1.66, 3.61)$ (in order from the first to fifth peaks) and $(d_1,d_2) = (9.43 \times 10^{-4}, 0.368)$ for ${\rm Im}\,\Delta \epsilon_\parallel$ and $(2.00, 0.00), (1.97, 0.423), (1.83, 1.76), (1.75, 2.71), (1.70, 3.46)$ and $(d_1,d_2) = (6.85 \times 10^{-4}, 0.292)$ for ${\rm Im}\,\Delta \epsilon_\perp$.  
}
\end{figure*}

Let us have a closer look at the oscillating peak structure; see Fig.~\ref{fig3}, where the peak locations and heights are plotted against the field strength $\bar{E}$.  Note that we identify the first peak as the positive sharp peak at $\omega/m=2$ in Fig.~\ref{fig2} and the second as the negative peak right after the first one around $\omega/m \approx 2.1\,\mathchar`-\,2.2 $, and successively define the $n$-th peaks.

We first discuss the peak locations; see the top of Fig.~\ref{fig3}.  The first peak appears precisely at the threshold band-gap energy $\omega = 2m$, regardless of the field strength $\bar{E}$.  The independence of $\bar{E}$ is simply because the first peak is caused by the threshold behavior of the free-field result $\epsilon_{ij}(\bar{E}=0)$ (\ref{eq::::8}).  In contrast, the subsequent peaks show monotonically increasing behaviors in $\bar{E}$ and can be fit well with a square-root function,
\begin{align}
	\frac{\omega}{m} = c_1 + c_2 \sqrt{\frac{e\bar{E}}{m^2} } \;. \label{gqewdqwedwq:73}
\end{align}  
See the caption of Fig.~\ref{fig3} for the best-fit parameters $(c_1, c_2)$.  The square-root $\bar{E}$ dependence reflects how the Dirac-sea structure changes with $\bar{E}$.  Let us make a rough semi-classical argument on this.  Due to the presence of the constant strong electric field, the electron wave-function in the Dirac sea would acquire an additional phase factor ${\rm e}^{\pm{\rm i}pz}  \to {\rm e}^{\pm {\rm i}\int {\rm d}z\,(p - e\bar{E}z )} \sim {\rm e}^{\mp {\rm i}e\bar{E} z^2}$.  The plus and minus signs, respectively, specify the moving direction of the Dirac-sea electrons heading to and receding from the gap.  Those two waves interfere with each other.  Thus, the resulting electron distribution is oscillating in the $z$ direction $\propto \cos (2e\bar{E} z^2)$, which has the period of $\Delta z \propto 1/\sqrt{\bar{E}}$.  This indicates that the electron distribution, when viewed as a function of the electron energy $\varepsilon$, has maxima with the period $\Delta \varepsilon = e\bar{E} \Delta z  \propto \sqrt{\bar{E}}$ because the electron potential energy is tied with the $z$ coordinate as $\varepsilon = -e\bar{E}z$.  The pair production is thus more likely to occur and develop a larger imaginary part when the probe frequency $\omega$ matches the energies $\varepsilon$ at which the distribution has maxima.  This means that the imaginary part has maxima with the period $\Delta \omega = \Delta \varepsilon \propto \sqrt{\bar{E}}$.  Therefore, the peak positions scale roughly as $ \propto \sqrt{\bar{E}}$, as observed in Fig.~\ref{fig3}.  Note that semiconductor calculations predict $\Delta \omega \propto \bar{E}^{3/2}$~\cite{PhysRev.130.549}.  The difference is stemming from the non-relativistic dispersion of solid-state materials.

The peak locations are less sensitive to the probe directions.  Indeed, the deviations between the peak locations of the longitudinal and transverse electric permittivities in Fig.~\ref{fig3} are just a few percent.  This is physically reasonable, as the peak locations are essentially determined by the modified Dirac-sea structure by the strong field, as explained above.  Therefore, the probe can only have a minor effect on the locations.

In contrast, the heights of the peaks are sensitive to the probe directions; see the bottom of Fig.~\ref{fig3}.  Indeed, the heights are directly related to the size of the dielectric energy loss, i.e., how many pairs are produced, which is determined by how they are excited by the probe and hence should be sensitive to the probe profile.  Figure~\ref{fig3} shows that the height of the first peak of the transverse electric permittivity ${\rm Im}\,\Delta \epsilon_\perp$ is significantly smaller than the longitudinal one ${\rm Im}\,\Delta \epsilon_\parallel$.  We can understand this as reminiscent of that a longitudinal probe enhances the particle production in the low-frequency regime more than a transverse one does [see Eq.~(\ref{eq::::a8})].  This is because the magnitude of the total electric field $|{\bm E}| = |\bar{\bm E} + {\bm {\mathcal E}}|$ is maximized (minimized) when $\bar{\bm E} \parallel {\bm {\mathcal E}}$ ($\bar{\bm E} \perp {\bm {\mathcal E}}$).  In contrast to the first peak, the tendency is found to be the opposite for higher peaks ($n\geq 3$): the peak heights of the transverse electric permittivity ${\rm Im}\,\Delta \epsilon_\perp$ are larger than the longitudinal one ${\rm Im}\,\Delta \epsilon_\parallel$.  This implies that it is easier with a transverse probe to observe the change of the electric permittivity in the high-frequency regime.

From an experimental viewpoint, the largest peaks, i.e., the first peaks (the red lines on the bottom of Fig.~\ref{fig3}) are of particular interest, as they may be the easiest to be accessed by experiments among all the other peaks.  To better understand the parameter dependence, we have tried fitting of the curves with elementary functions and eventually found that a power-function,  
\begin{align}
	{\rm Im}\,\Delta \epsilon_{ij}  = d_1 \left( \frac{e\bar{E}}{m^2} \right)^{d_2}, \label{gqerqe:74}
\end{align}
gives the best fit; see the caption of Fig.~\ref{fig3} for the best fitting parameters $(d_1,d_2)$.  Note that the height is free from the non-perturbative exponential suppression $\propto {\rm e}^{-\pi m^2/e\bar{E}}$, which is effective only in the low-frequency regime (or the semi-classical regime where the semi-classical approximations such as the worldline instanton method~\cite{Affleck:1981bma, Dunne:2005sx, Dunne:2006st, Dunne:2006ur, Taya:2020dco} can be applied).  The exponent $d_2$ roughly equals $1/3$, which agrees with the value predicted in semiconductor calculations~\cite{PhysRev.140.A263}.  This compact formula (\ref{gqerqe:74}) enables us to discuss the weak-field regime $e\bar{E}/m^2 \ll 1$ rather easily, as the direct numerical evaluation of our formula (\ref{g2e;::15ffefq}) becomes difficult for small field strengths [this is because Eq.~(\ref{g2e;::15ffefq}) contains the non-perturbative factor ${\rm e}^{-\pi m^2/e\bar{E}}$, which becomes extremely small for $e\bar{E}/m^2 \to 0$, and therefore it becomes difficult to assure the numerical precision].  It is also useful for making experimental predictions (see Sec.~\ref{sec:5}), for which weak fields are more relevant owing to the current unavailability of strong fields.

Finally, we point out another important feature of Fig.~\ref{fig2}: the imaginary part of the electric permittivity is non-vanishing even at $\omega = 0^+$\footnote{The oddness of the imaginary part (\ref{egqefsreqr:18}) indicates that it is vanishing precisely at $\omega = 0$.  This is compatible with the finiteness at $\omega = 0^+$, as $\omega = 0^+$ means $\omega$ is infinitely close to zero and does not mean $\omega = 0$, though it is inevitable to have a discontinuity at $\omega = 0$.  }.  This is a non-perturbative strong-field effect $\propto {\rm e}^{-\pi m^2/e\bar{E}}$, which is evident from Eq.~(\ref{eq:::::::a8}).  Physically, the pair production occurs, and therefore non-zero imaginary part develops, even with a zero-frequency probe due to the additional energy supply by the strong field.  Note that the imaginary part is vanishing exactly in the weak-field limit, where all the non-perturbative factors are approximated to be zero.  The non-zero imaginary part at $\omega = 0^+$ affects the $\omega = 0^+$ behavior of the real part through the Kramers-Kr\"{o}nig relation, and thereby makes it deviating from the conventional result for the weak and low-frequency limit; see Sec.~\ref{sec:4b}.

\subsection{Real part ${\rm Im}\,\Delta \epsilon_{ij}$} \label{sec:4b}

\begin{figure*}[!t]
\includegraphics[clip, width=0.495\textwidth]{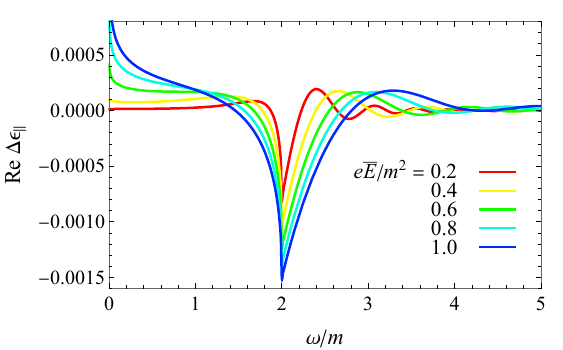}
\includegraphics[clip, width=0.495\textwidth]{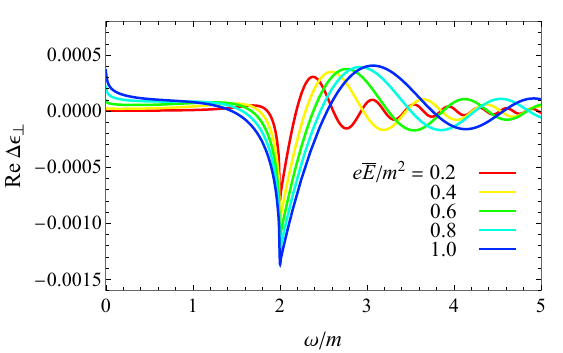}
\caption{\label{fig4} A similar plot to Fig.~\ref{fig2} but for the real part.  Namely, the change of the real part of the longitudinal and transverse electric permittivities, ${\rm Re}\,\Delta\epsilon_\parallel$ (left) and ${\rm Re}\,\Delta\epsilon_\perp$ (right), are shown as functions of the probe frequency $\omega$. 
}
\end{figure*}

Figure~\ref{fig4} shows the frequency $\omega$ dependence of the real part ${\rm Re}\,\Delta\epsilon_{ij}$ (\ref{gqerqq:40-aa-}).  As the real and imaginary part are closely related with each other through the Kramers-Kr\"{o}nig relation (\ref{gqerqq:40-aa-}), the real part also exhibits a similar oscillating structure.  Note that very similar oscillating behaviors have also been observed in various solid-state materials (e.g., in Ge~\cite{PhysRevLett.14.138}).

\begin{figure*}[!t]
\hspace*{8.2mm}\includegraphics[clip, width=0.445\textwidth]{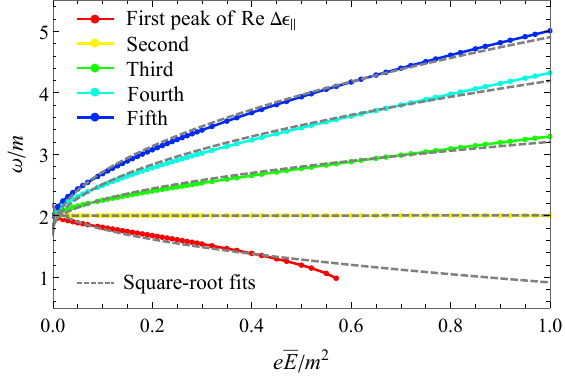}
\hspace*{8.8mm}\includegraphics[clip, width=0.445\textwidth]{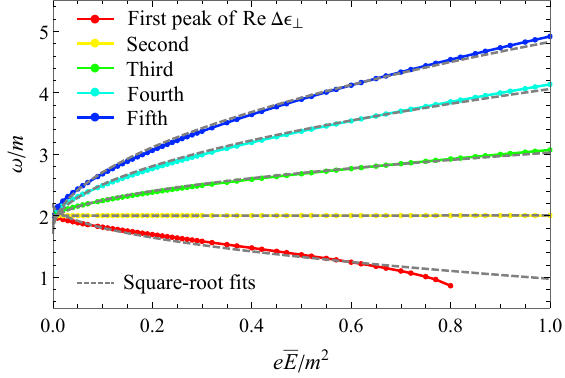} \\
\includegraphics[clip, width=0.495\textwidth]{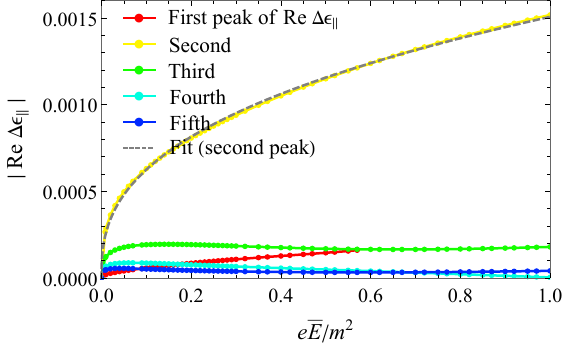}
\includegraphics[clip, width=0.495\textwidth]{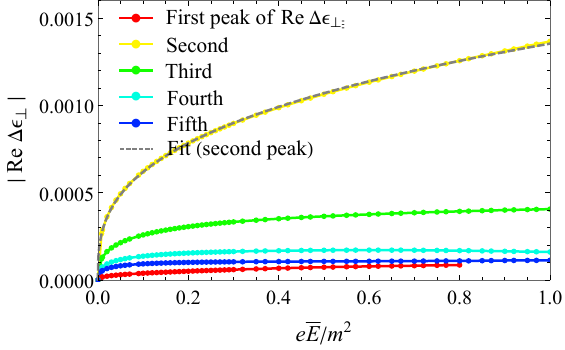}
\caption{\label{fig5} A similar plot to Fig.~\ref{fig3} but for the real part.  Namely, the peak locations (with respect to $\omega$) (top) and heights (bottom) of the change of the electric permittivities, ${\rm Re}\,\Delta\epsilon_\parallel$ (left) and ${\rm Re}\,\Delta\epsilon_\perp$ (right), are shown as functions of the strength of the strong field $e\bar{E}/m^2$.  The gray dashed lines show fitting results of the curves, i.e., square-root fits $\omega/m = c_1 + c_2 \sqrt{e\bar{E}/m^2}$ for each peak location on the top and a power-function fit ${\rm Re}\,\Delta \epsilon_{ij} = d_1 (e\bar{E}/m^2)^{d_2}$ for the heights of the second peaks, which are the largest among the peaks, on the bottom.  The best-fit parameters $(c_1, c_2)$ are found to be $(2.18, -1.27), (2.00, 0.00), (1.82, 1.38), (1.73, 2.46), (1.70, 3.21)$ (in order from the first to fifth peaks) and $(d_1,d_2) = (-1.51 \times 10^{-3}, 0.382)$ for ${\rm Re}\,\Delta \epsilon_\parallel$ and $(2.19, -1.21), (2.00, 0.00), (1.90, 1.13), (1.79, 2.27), (1.72, 3.10)$ and $(d_1,d_2) = (-1.35 \times 10^{-3}, 0.340)$ for ${\rm Re}\,\Delta \epsilon_\perp$.  
}
\end{figure*}

We discuss details about the oscillating structure; see Fig.~\ref{fig5}.  Note that we have identified the first peak as the small positive bump just before the threshold band-gap energy $\omega < 2m$ [e.g, $\omega \approx 1.9$ in the red lines ($e\bar{E}/m^2=0.2$) of Fig.~\ref{fig4}] and the second peak as the largest negative sharp peak at $\omega \approx 2m$, and successively define the higher peaks.  As we shall discuss later, the first peak can be absent for strong fields, for which the real part diverges logarithmically at $\omega \approx 0$ and then the first peak can be hidden under the divergence.  This is the reason why the red curves in Fig.~\ref{fig5} disappears at around $e\bar{E}/m^2 \approx 0.57$ and $\approx 0.80$ for $\epsilon_{\parallel}$ and $\epsilon_{\perp}$, respectively.

As the oscillating structure is reminiscent of that in the imaginary part, the peak locations can be fit well by the square-root function (\ref{gqewdqwedwq:73}); see the top of Fig.~\ref{fig5}.  Nevertheless, although it can be fit by the same function (\ref{gqewdqwedwq:73}), the fitting parameters $(c_1,c_2)$ and thus the values of the locations are different.  The peak locations of the real part are roughly in the middle of those for the imaginary part [e.g., the location of the third peak of the real part at $e\bar{E}/m^2 = 1.0$ is $\omega/m \approx 3.1$, which is just around the middle of the second ($\omega/m \approx 2.5$) and third ($\omega/m \approx 3.9$) peaks of the imaginary part].  From the viewpoint of the Kramers-Kr\"{o}nig analysis, this is because the Hilbert transformation is a transformation such that it splits a peak of the integrand into two.  To understand this, let us consider, as an example, the Lorentz model of dielectrics~\cite{lorentz1916theory}, which is widely used in the phenomenological modeling of the electric permittivity in solid-state physics.  The imaginary part in this model is
\begin{align}
	{\rm Im}\,\Delta \epsilon^{\rm Lorentz} := \omega_{\rm p}^2 \frac{\Gamma \omega}{(\omega^2-\omega_0^2)^2 + (\Gamma \omega)^2} \;, \label{fqefreqrq:76}
\end{align}
where $\omega_{\rm p}^2$ is some constant (identified as the plasma frequency in the original context).  The imaginary part ${\rm Im}\,\Delta \epsilon^{\rm Lorentz}$ has a single positive peak at $\omega = \omega_0$ with the width $\Gamma$.  Now, we Hilbert-transform Eq.~(\ref{fqefreqrq:76}) to see how the Hilbert transformation acts on a peak.  This is exactly doable and yields
\begin{align}
	{\rm Re}\,\Delta \epsilon^{\rm Lorentz} = -\omega_p^2 \frac{\omega^2-\omega_0^2}{(\omega^2-\omega_0^2)^2 + (\Gamma \omega)^2} \;. \label{gqereqrq:77}
\end{align}
Differentiating with respect to $\omega$, we immediately understand that the real part (\ref{gqereqrq:77}) has two peaks around $\omega = \omega_0$ separated by the distance $\Gamma$; namely, a positive peak at $\omega = \sqrt{\omega_0(\omega_0 -\Gamma)} \approx \omega_0 - \Gamma/2$ and a negative peak at $\omega = \sqrt{\omega_0(\omega_0 + \Gamma)} \approx \omega_0 + \Gamma/2$.  This confirms our statement that the Hilbert transformation splits a peak into two.  Now, we consider applying the Lorentz-model argument to our problem.  In general, including our electric permittivity, the imaginary part of the electric permittivity can have several peaks, which can be modeled by superimposing the Lorentz models (\ref{fqefreqrq:76}) with various peak profiles.  We can, thus, understand the peaks of the real part in Fig.~\ref{fig5} as a collection of those of the imaginary part in Fig.~\ref{fig3} after the splitting by the Hilbert transformation.  For example, the first positive peak of the imaginary part at $\omega/m = 2$ splits into two, one of which gives a positive peak below the threshold band-gap energy and the other is a negative peak above it\footnote{The first peak of the imaginary part is very sharp and skewed in our case, and therefore the application of the Lorentz model is actually a bit naive.  In fact, the second peak of the real part is always sticked at $\omega/m \approx 2$ and does not clearly show the positive shift from the original peak position of the imaginary part.  Nonetheless, it is still useful to get an intuition of the peak locations of the real part, as we explain in the main text.  }, and those peaks are separated roughly by the peak width of the imaginary part.  The location of the peak of the real part below the band-gap decreases its value as the field strength increases.  This is because, as the field strength becomes larger, the width becomes larger (see Fig.~\ref{fig2}), which is essentially because the Schwinger effect at $\omega = 0$ creates more particles and accordingly the exponential tail below the band-gap develops more.  A similar argument applies to the second and higher peaks: the peak locations of the real part are basically shifted from those of the imaginary part by the peak widths.  Consequently, the peak locations of the real part are roughly in the middle of those for the imaginary part.

We turn to discuss the peak heights; see the bottom of Fig.~\ref{fig5}.  The basic features are qualitatively similar to the imaginary case because the real part is tied with the imaginary part through the Kramers-Kr\"{o}nig relation (\ref{gqerqq:40-aa-}).  Namely, (i) the heights are more sensitive to the probe direction than the locations are, i.e., the real part also exhibits the birefringent nature of the vacuum in strong fields.  (ii) Except for the largest second peak, the heights are relatively larger in the transverse electric permittivity than in the longitudinal one.  The heights of the second peaks are roughly comparable between the longitudinal and transverse cases; though the transverse one is slightly smaller ($\approx 10\;\%$) than the longitudinal one, which is reasonable as the largest peak of the imaginary part (which gives the dominant contribution to the second peak of the imaginary part, as we have explained in the last paragraph) of the transverse electric permittivity is smaller than the longitudinal one.  (iii) The largest second peak can be fit well by the power-function (\ref{gqerqe:74}); see the caption of Fig.~\ref{fig5} for the best fitting parameters $(d_1,d_2)$.  Compared to the largest peak of the imaginary part (see Fig.~\ref{fig3}), the values of the exponent $d_2$ are somewhat increased ($\approx 10\;\%$) but stay around $1/3$, indicating that the $e\bar{E}/m^2$ dependence is essentially unchanged.  Contrary, the overall factor $d_1$ is significantly increased ($\approx 100\;\%$) and thus the real part has larger peak heights than the imaginary part.

Next, let us discuss the low-frequency behavior of the real part of the electric permittivity.  We have found it logarithmically divergent; see Fig.~\ref{fig4}.  This is originating from the finiteness of the imaginary part at $\omega = 0^+$ [see Eq.~(\ref{eq:::::::a8}) and Fig.~\ref{fig2}], which yields from the Kramers-Kr\"{o}nig relation (\ref{gqerqq:40-aa-}) that 
\begin{align}
	&{\rm Re}\,\Delta\epsilon_{ij}(\omega) 
	\xrightarrow{\omega \to 0} \frac{-2\,{\rm Im}\,\Delta \epsilon_{ij}(0)}{\pi} {\rm ln}\,\omega\, \label{gqefreqr:80} \\
	&= \left\{ \begin{array}{ll} 
		\displaystyle \frac{-2\alpha}{\pi^2} {\rm e}^{-\pi\frac{m^2}{e\bar{E}}} \left( \frac{1}{\pi} + \frac{1}{2} \frac{m^2}{e\bar{E}} \right) {\rm ln}\,\omega & {\rm for}\ \perp \;, \\[10pt]
		\displaystyle \frac{-2\alpha}{\pi^2} {\rm e}^{-\pi\frac{m^2}{e\bar{E}}} \left( \frac{1}{\pi} + \frac{m^2}{e\bar{E}} + \frac{\pi}{2} \left( \frac{m^2}{e\bar{E}} \right)^2 \right) {\rm ln}\,\omega & {\rm for}\ \parallel \;.
  		\end{array} \right. \nonumber
\end{align}
Equation~(\ref{gqefreqr:80}) clearly shows that the logarithmic divergence is a non-perturbative strong-field effect $\propto {\rm e}^{-\pi m^2/e\bar{E}} $, and hence is dismissed in the weak-field calculations.

\begin{figure*}[!t]
\includegraphics[clip, width=0.495\textwidth]{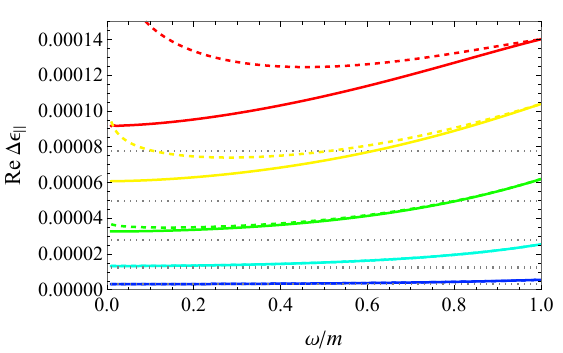}
\includegraphics[clip, width=0.495\textwidth]{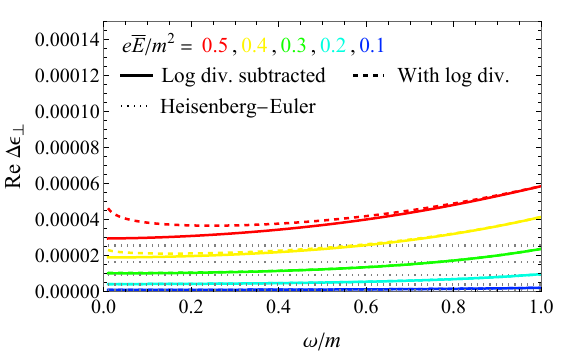}
\caption{\label{fig6} A comparison between our results (thick and dashed color lines) and the Heisenberg-Euler result (\ref{fqefedada:74}) (dotted gray lines), which is valid in the weak and zero-frequency limit.  The thick and dashed lines represent the real part with and without subtracting the logarithmic divergence, Eqs.~(\ref{gqerqedq:80}) and (\ref{gqerqq:40-aa-}), respectively.  
}
\end{figure*}

To get a better understanding of the logarithmic divergence and also the relation between the known result in the literature and ours, we made Fig.~\ref{fig6}, where the known result based on the Heisenberg-Euler effective Lagrangian [see Eq.~(\ref{fqefedada:74}) below] is compared with ours (\ref{gqerqq:40-aa-}) and that with subtracting the logarithmic divergence (\ref{gqefreqr:80}) by hand,  
\begin{align}
	&{\rm Re}\,\Delta\epsilon_{ij}^{\rm log\ sub.}  \label{gqerqedq:80}\\
	&:= \left\{ \begin{array}{ll} 
		\displaystyle {\rm Re}\,\Delta\epsilon_{\perp} &\\[3pt]  \displaystyle  - \frac{-2\alpha}{\pi^2} {\rm e}^{-\pi\frac{m^2}{e\bar{E}}} \left( \frac{1}{\pi} + \frac{1}{2} \frac{m^2}{e\bar{E}} \right) {\rm ln}\,\omega & {\rm for}\ \perp \;, \\[25pt]
		\displaystyle {\rm Re}\,\Delta\epsilon_{\parallel} &\\[3pt] \displaystyle  - \frac{-2\alpha}{\pi^2} {\rm e}^{-\pi\frac{m^2}{e\bar{E}}} \left( \frac{1}{\pi} + \frac{m^2}{e\bar{E}} + \frac{\pi}{2} \left( \frac{m^2}{e\bar{E}} \right)^2 \right) {\rm ln}\,\omega & {\rm for}\ \parallel \;.
  		\end{array} \right. \nonumber
\end{align}
As shown in Fig.~\ref{fig6}, the subtracted number (\ref{gqerqedq:80}) is finite at $\omega \to 0$.  It also shows that the subtraction does not change the value significantly for weak fields, meaning that the logarithmic divergence can be significant only for strong fields.

In the literature, the electric permittivity has been typically calculated with the Heisenberg-Euler effective Lagrangian and found to be (see Appendix~\ref{app:C} for the derivation)~\cite{PhysRev.135.B1279, Baier:1967zzc}
\begin{align}
	{\rm Re}\,\Delta\epsilon^{\rm HE}_{ij}  
	= \left\{ \begin{array}{ll} 
		\displaystyle +  \frac{\alpha}{\pi} \frac{2}{45}  \left( \frac{e\bar{E}}{m^2} \right)^{2}  & {\rm for}\ \perp \;, \\[10pt]
		\displaystyle +  \frac{\alpha}{\pi}  \frac{2}{15} \left( \frac{e\bar{E}}{m^2} \right)^{2} & {\rm for}\ \parallel \;,
  		\end{array} \right. \label{fqefedada:74} 
\end{align}
which is valid in the weak and slow limit $\bar{E}, \omega \to 0$.  Note that the imaginary part is strictly zero in this approach.  Figure~\ref{fig6} shows that our result, which is valid even to strong fields and finite frequencies, coincides with the Heisenberg-Euler result (\ref{fqefedada:74}) at and only at $\bar{E}, \omega \to 0$ (the same applies to the imaginary part; see Fig.~\ref{fig2}, which shows that our results go to zero at $\bar{E}, \omega \to 0$).  The agreement becomes better if we drop the logarithmic divergence, as this is a part of the weak-field approximation made in the Heisenberg-Euler approach.  Our results deviate from Heisenberg-Euler one for strong fields $e\bar{E}/m^2 \gtrsim 0.2$ at $\omega \approx 0$.  Meanwhile, for weak fields below this value, we observe good agreements between the two for wide values of the frequency below the threshold band-gap energy $\omega/m \lesssim 1$.  Thus, we conclude that our result is consistent with the conventional Heisenberg-Euler result in the weak and slow limit, while significant deviations appear for strong or fast fields, for which the Heisenberg-Euler result (\ref{fqefedada:74}) is invalid and ours (\ref{gqerqq:40-aa-}) must be used instead.

\section{Summary and discussion} \label{sec:5}

In summary, we have revisited the change of the electric permittivity, as a response of the vacuum against a probe electric wave, in the presence and absence of a strong constant electric field.  As described in Sec.~\ref{sec:3}, our calculation is based on a linear-response theory in the strong electric field, which we have formulated using the non-equilibrium in-in formalism and the Furry-picture perturbation theory.  Our in-in formulation is manifestly causal, which naturally yields the Kramers-Kr\"{o}nig relation.  By subtracting the infrared divergence appropriately, we have first calculated the imaginary part of the change of the electric permittivity (\ref{g2e;::15ffefq}) and then made use of the Kramers-Kr\"{o}nig relation to numerically obtain the real part (\ref{gqerqq:40-aa-}) and thereby the whole electric permittivity.  We have also established a clear relation between the electric permittivity and the number of pairs produced by the dynamically-assisted Schwinger effect (\ref{eq;vwefewqf50}).  Our approach is beyond the conventional weak-field and zero-frequency-probe limit, and has revealed intriguing features for strong fields and/or probes with finite frequencies; see Sec.~\ref{sec:4}.  In particular, we have found (for the first time in the context of strong-field QED to the best of our knowledge) that both the real and imaginary parts of the electric permittivity exhibit characteristic oscillating patterns with respect to the probe frequency.  This is quite analogous to what has been observed in semiconductor experiments in the context of the Franz-Keldysh effect and electroreflectance.  The oscillating structure is directly related to the change of the QED-vacuum structure (cf. the square-root $\bar{E}$ dependence of the peak locations).  This implies that the measurement of the electric permittivity serves as a probe to quantitatively diagnose the QED vacuum.  Another notable feature is that the low-frequency behavior is modified due to a non-perturbative strong-field effect.  This modification is suppressed strongly by the field strength, ${\rm e}^{-\pi m^2/e\bar{E}}$, and therefore is just a minor effect for weak fields but can have significant impacts for strong fields such as the logarithmic divergence in the real part.

Let us discuss possible experimental and phenomenological implications.  As mentioned in Introduction, strong electromagnetic fields can be realized in extreme physical systems such as heavy-ion collisions and compact stars and also with recent/near-future high-power lasers under laboratory conditions.  Unfortunately, the direct application of our results to the extreme physical systems is premature, since the electromagnetic-field configurations realized there are very different to our simple constant-electric-field configuration (\ref{eq:2}).  For example, the electromagnetic fields in heavy-ion collisions have extremely small spacetime volume, for which our constant-field approximation cannot be applied.  Compact stars such as magnetars and charged black-holes may have strong electric fields, but other strong-field effects including magnetic and/or gravitational ones are more significant.  Therefore, our present calculation with the simple configuration (\ref{eq:2}) is premature to discuss anything concrete in those extreme physical systems, which we leave as future work.  Nonetheless, we expect that the essence of the physics should not change depending on the field configurations.  That is, our basic finding that the change of the QED vacuum structure by a strong field leads to intriguing frequency- and polarization-dependent changes in the electric permittivity, or the refractive index in general, should be valid.  Such a change would be the most significant at around the gap energy $\omega \sim m$.  It is interesting and worthwhile to try observing the change with, e.g., the recent and future photon- and/or di-lepton- polarization measurements in heavy-ion collisions and astronomical X-ray observations (e.g., with IXPE and XL-Calibur).

Let us next discuss implications with high-power lasers in a quantitative manner.  High-power lasers are advantageous than the extreme physical systems in that the field configuration can be designed rather easily and thus in principle our simple field configuration (\ref{eq:2}) can be realized.  It would also be advantageous in that experimental noise is suppressed and can be controlled.  Therefore, it is more motivated to consider the case of high-power lasers than the extreme physical systems.  In short, however,  it seems difficult to observe our frequency-dependent change of the electric permittivity within the current experimental technology.  The most difficult part (but at the same time {\it only} the problem) is that extremely fast coherent-light sources of the order of the zeptosecond, $\omega \gtrsim m = 1.24 \times 10^{20}\;{\rm Hz}$ (for electron), are needed as probes.  Such a zeptosecond light source is still unavailable\footnote{Instead of using coherent lights, it may be possible to use high-energy incoherent photons as the fast probe field such as bremsstrahlung photons emitted from and highly-Lorentz-boosted Coulomb field of high-energy electrons (or charged particles such as proton or heavy ions).  This is an interesting possibility, but our setup (\ref{eq:2}) is too idealistic for such a situation (we need to consider {\it incoherent} photons rather than the coherent ones, finite photon momentum ${\bm k} \neq {\bm 0}$, and so on), and also we need to re-formulate our theory to extract the information of the electric permittivity from incoherent photons.  Thus, we do not make any quantitative discussion for this possibility.  }, though there have been significant developments in ultrafast lasers in the attosecond regime in the last decade~\cite{RevModPhys.81.163, Popmintchev2010, Li2020, Midorikawa_2011} and exist various proposals to achieve the zeptosecond order (e.g., Ref.~\cite{PhysRevLett.93.115002, klaiber2007zeptosecond, PhysRevResearch.4.L032038}).  To proceed, let us suppose anyway that such an extremely fast light source is available and discuss what we can say then.  The easiest to be observed is the largest peak at $\omega \approx 2m$.  The peak height can be approximated well by Eq.~(\ref{gqerqe:74}).  In physical units, it reads
\begin{align}
\begin{split}
	{\rm Im}\,\Delta \epsilon_\parallel  
		&= (1.1 \times 10^{-4}) \times \left( \frac{I}{1 \times 10^{23}\;{\rm W/cm}^2} \right)^{0.15}  \;, \\
	{\rm Im}\,\Delta \epsilon_\perp 
		&= (0.98 \times 10^{-4}) \times \left( \frac{I}{1 \times 10^{23}\;{\rm W/cm}^2} \right)^{0.18} \;, \label{gqereqreq:3311}
\end{split}
\end{align}
and
\begin{align}
\begin{split}
	{\rm Re}\,\Delta \epsilon_\parallel  
		&\approx (-1.4 \times 10^{-4}) \times \left( \frac{I}{1 \times 10^{23}\;{\rm W/cm}^2} \right)^{0.19} \;, \\
	{\rm Re}\,\Delta \epsilon_\perp  
		&\approx (-1.7 \times 10^{-4}) \times \left( \frac{I}{1 \times 10^{23}\;{\rm W/cm}^2} \right)^{0.17} \;, \label{gqereqreq:3311--}
\end{split}
\end{align}
where $I := \bar{E}^2/2$ is the intensity of a tightly focused laser pulse.  As an example, let us consider the state-of-art strongest pulse with a 4-PW laser~\cite{Yoon:21}: $I = 1 \times 10^{23}\;{\rm W/cm^2}$ (or $\bar{E} = 9\times 10^{14}\;{\rm V/m}$) with $\sigma = 1\;{\rm \mu m}$ and $\tau = 20\;{\rm fs}$ being the spot size and the time duration of the pulse, respectively.  The estimates, (\ref{gqereqreq:3311}) and (\ref{gqereqreq:3311--}), indicate that the order of the change is $\Delta \epsilon_{ij} = 1 \times 10^{-2}\;\%$ and the difference between the longitudinal and transverse ones (i.e., the signature of the vacuum birefringence) is $|\Delta \epsilon_{\parallel} - \Delta \epsilon_{\perp}|= 2 \times 10^{-3}\;\%$.  Those are not large numbers but are significantly larger compared to the weak-field and low-frequency limit (\ref{fqefedada:74}), $\Delta \epsilon^{\rm HE}_{ij} = {\mathcal O}(10^{-5}\;\%)$.  We emphasize that the estimates, (\ref{gqereqreq:3311}) and (\ref{gqereqreq:3311--}), depend on the intensity very weakly $\Delta \epsilon_{ij} \propto I^{d_2/2} \approx I^{1/6}$.  This means that sizable signals can remain even for weak fields.  Namely, to lower the magnitude of the change by one order, we need to change the intensity by about six orders.  For example, even for a GW laser, we still have a sizable magnitude of the changes, $\Delta \epsilon_{ij} = 1 \times 10^{-3}\;\%$ and $|\Delta \epsilon_{\parallel} - \Delta \epsilon_{\perp}|= 2 \times 10^{-4}\;\%$.  Those changes may be measured directly with the standard technique of dielectric spectroscopy (see, e.g., Ref.~\cite{kremer2002broadband}), or indirectly through measurement of other observables.  One possible observable is the change of the speed of light in the vacuum $c$, which is related to the electric permittivity as $c = 1/\sqrt{\epsilon \mu}$ (with $\mu$ being the magnetic permeability) and therefore the change of $\epsilon$ immediately means that of $c$.  Our result is, however, premature to discuss anything quantitative about the change of the speed of light (e.g., need to calculate the change of the magnetic permeability, to include photon momentum ${\bm k} \neq {\bm 0}$, and so on); nonetheless, we may naively expect a similar order of the change appears in the speed of light as well.  If so, we have $\Delta c = {\mathcal O}(10^{-2}\;\%)$ and $|\Delta c_{\parallel} - \Delta c_{\perp}|= {\mathcal O}(10^{-3}\;\%)$ for photons around the threshold band-gap energy $\omega \approx 2m$, which may be tested with, e.g., the photon polarimeter (e.g., PVLAS~\cite{Ejlli:2020yhk}) and the Michelson-Morley interferometer~\cite{Michelson333}.  Another interesting observable is the number of pairs produced by the dynamically-assisted Schwinger effect, which is related to the imaginary part through Eq.~(\ref{eq;vwefewqf50}).  In physical units, it reads\footnote{To be precise, the left-hand side of Eq.~(\ref{tqrwr2:llll}) should be understood as $N({\mathcal E} \neq 0) - N({\mathcal E} = 0)$, i.e., the number of pairs produced solely by the probe field.  For weak fields $e\bar{E} \ll m^2$, which are relevant for the current experimental technology, the Schwinger contribution without the probe is negligible $N({\mathcal E} = 0) \approx 0$.  Therefore, $N({\mathcal E} \neq 0) - N({\mathcal E} = 0) \approx N({\mathcal E} \neq 0)$, i.e., we do not need to carefully distinguish the total number of produced pairs and those solely by the probe.  }
\begin{align}
	N	&\approx \frac{1}{6 \alpha} m^4 \sigma^3 \tau \left[  \left( \frac{ e{\mathcal E}_\perp }{m^2} \right)^2 {\rm Im}\;\Delta \epsilon_\perp  +  \left( \frac{e{\mathcal E}_z}{m^2} \right)^2 {\rm Im}\;\Delta \epsilon_\parallel  \right] \nonumber\\
		&\approx 0.20 \times \left(\frac{\sigma}{\mu {\rm m}}\right)^3 \left( \frac{\tau}{{\rm fs}} \right) \nonumber\\
			&\quad \times \left[  0.98 \times \left( \frac{ {\mathcal E}_\perp }{ 1\times 10^{15}\;{\rm V/m} }\right)^2 \left( \frac{I}{1 \times 10^{23}\;{\rm W/cm}^2} \right)^{0.18}  \right. \nonumber\\
				&\quad\quad \left. +  1.1 \times \left( \frac{ {\mathcal E}_z }{ 1\times 10^{15}\;{\rm V/m} } \right)^2 \left( \frac{I}{1 \times 10^{23}\;{\rm W/cm}^2} \right)^{0.15}  \right]  \label{tqrwr2:llll}
\end{align} 
per a single shot of a laser pulse.  Suppose we have pulses with a repetition rate of $f = 0.1\;{\rm Hz}$ and take the probe-field strength to be ${\mathcal E} = 1 \times 10^{12}\;{\rm V/m}$ (corresponding to the GW regime with the same focusing parameters).  Then, the total number of pairs produced per unit time with the 4-PW-laser configuration is the order of $N/{\rm sec} \approx (4 \times 10^{-5}) \times f \approx 4 \times 10^{-6}$, indicating that a pair is produced per three days.  For the difference between the longitudinal and transverse probes, $(N_{\parallel} - N_{\perp}) /{\rm sec} \approx (6 \times 10^{-6}) \times f \approx 6 \times 10^{-7}$, it roughly requires three weeks to observe.  As was the case in $\Delta \epsilon_{ij}$, the number of pairs is also suppressed weakly in $I$ as $N \appropto I^{1/6}$ and therefore even for weak fields we still have chance to observe the pair production and thereby the change of the electric permittivity.  Note that for a fixed peak power $P$, the intensity scales as $I \propto P/\sigma^2\ \Rightarrow \ N \appropto I^{1/6} \sigma^3 \propto \sigma^{8/3}$.  Thus, a looser focusing is more beneficial for the pair production (e.g., for $\sigma = 10\;\mu{\rm m}$, the rate $N/{\rm sec}$ is enhanced by the factor of about 500, $N/{\rm sec} \approx 2 \times 10^{-3}$, which means about 200 pairs are produced in a day).  We, thus, conclude this discussion that the observation of our predicted change of the electric permittivity is not easy at the moment but should be feasible once a zeptosecond light source becomes available.

We conclude this paper by again emphasizing the usefulness of the analogy between strong-field QED and semiconductor physics (and also other physical systems).  Although there are subtle differences between semiconductor and QED [semiconductor has smaller ${\mathcal O}(1\;{\rm eV})$ mass gap, the (effective) speed of light deviating from the vacuum one, impurities, lattice structure, dimensionality, etc; see, e.g., Ref.~\cite{Linder:2017gxp}], they essentially have the same band structure and hence respond against electromagnetic fields in a very similar manner.  Previously, the analogy between the dynamically-assisted Schwinger effect in QED and the Franz-Keldysh effect has been discussed and was successful~\cite{Taya:2018eng, Torgrimsson:2018xdf}.  Motivated by this success, we in this paper have pursued the QED analog of the electroreflectance in semiconductor physics and succeeded in finding the novel behaviors in QED (e.g., the oscillation in the high-frequency regime) and the useful relation connecting the Schwinger effect and the electric permittivity (\ref{eq;vwefewqf50}).  This clearly demonstrates that importing the wisdom of semiconductor physics is useful in developing a better understanding of strong-field QED.  Although we have mainly focused on {\it importing} semiconductor knowledge to QED, the converse (i.e., {\it export}) should also be useful.  For example, topological insulators can effectively be described in terms of the Dirac equation~\cite{SHEN_2011, shen2013topological}.  Relativistic dispersion is realized in, e.g., graphene~\cite{RevModPhys.81.109} and Weyl/Dirac semimetal~\cite{RevModPhys.90.015001}.  In those materials, our relativistic QED calculation is more relevant than the conventional non-relativistic semiconductor one.  Our results indicate that, although the qualitative features remain the same, the quantitative ones must change, e.g., the different scaling of the peak locations $\Delta \omega \propto \sqrt{\bar{E}}$ for relativistic and $\propto \bar{E}^{3/2}$ for non-relativistic (see Figs.~\ref{fig3} and \ref{fig5}).  To make a more realistic prediction to condensed-matter experiments, however, it is crucially important to carefully take into account the matter properties and therefore we do not make any further predictions here, which we leave as future work.  Finally, let us mention that the analogy may also be extended to other physical systems such as in the presence of strong color fields as realized in the early-stage dynamics of heavy-ion collisions (e.g., glasma~\cite{Lappi:2006fp}) and gravitational fields in the Universe.  As we have learned from the analogy, the change of the vacuum structure is the essence to modify vacuum response functions.  Such a change can equally be induced by strong fields other than electromagnetic one, and therefore it is natural to expect similar modifications to vacuum response functions.  It is an interesting topic to study such cases, to which our in-in formalism can be generalized straightforwardly, and discuss possible phenomenological consequences, e.g., modification of jets in heavy-ion collisions by the strong color fields and vacuum birefringence in curved spacetime.

\section*{Acknowledgments}\ HT is supported by JSPS KAKENHI under grant No. 22K14045 and the RIKEN special postdoctoral researcher program.  HT is grateful to the Nonequilibrium working group at RIKEN Interdisciplinary Theoretical and Mathematical Sciences and also to the 79th Fujihara Seminar (Prospects for High-Field Science 2023; PHFS2023) and the participants there for fruitful discussions.

\appendix

\section{Details of the evaluations} \label{AppendixA}

In this Appendix, we describe some calculation details of the imaginary part ${\rm Im}\,\epsilon_{ij}$ (\ref{erqasdasda:36}) and the number of pairs produced by the dynamically-assisted Schwinger effect $N$ (\ref{eq:13}).  The essential part of the calculations is the evaluation of the Fourier transformations of bi-spinor products of the mode functions $\psi^{\rm as}_{\pm,{\bm p},s}$ (${\rm as}= {\rm in},{\rm out}$).  We begin with reminding the exact forms of the mode functions $\psi^{\rm as}_{\pm,{\bm p},s}$ in Appendix~\ref{app:Dirac} and then perform the Fourier transformations for the imaginary part and the number in Appendices~\ref{app:A} and \ref{app:B}, respectively.

\subsection{Solution of the Dirac equation in a constant electric field} \label{app:Dirac}

The analytical expressions of the mode functions $\psi^{\rm as}_{\pm,{\bm p},s}$ are known~\cite{Tanji:2008ku, Taya:2018eng, Huang:2019uhf}.  To display them, we decompose the mode functions $\psi^{\rm as}_{\pm,{\bm p},s}$ in the spinor space as
\begin{align}
	\psi_{+,{\bm p},s}^{\rm as}(t) &= \left[ A_{{\bm p}}^{\rm as}(t) + B_{{\bm p}}^{\rm as}(t) \gamma^0 \frac{ m + {\bm \gamma}_{\perp} \cdot {\bm p}_{\perp}}{\sqrt{m^2 + {\bm p}_{\perp}^2} } \right] \Gamma_s  \;, \nonumber\\
	\psi_{-,{\bm p},s}^{\rm as}(t) &= \left[ B_{{\bm p}}^{{\rm as}*}(t) - A_{{\bm p}}^{{\rm as}*}(t) \gamma^0 \frac{ m + {\bm \gamma}_{\perp} \cdot {\bm p}_{\perp}}{\sqrt{m^2 + {\bm p}_{\perp}^2} } \right] \Gamma_s \;.   \label{eqd31}
\end{align}
Substituting this spinor decomposition into the mode equation (\ref{fqewdasx;30}), it can be shown that the scalar functions $ A_{{\bm p}}^{\rm as},  B_{{\bm p}}^{\rm as}$ obey the Weber equation, which is analytically solvable.  For the boundary conditions (\ref{eq-10}) and (\ref{aaaeq-10}), the solutions read
\begin{align}
	&\left\{\begin{array}{l}
		A^{\rm in}_{{\bm p}}(t) =  {\rm e}^{-\frac{i\pi}{8}} {\rm e}^{- \frac{\pi a_{\bm p}}{4}} \sqrt{a_{\bm p}} D_{{\rm i} a_{\bm p} -1} \left(-{\rm e}^{-\frac{{\rm i}\pi}{4}} \xi_{\bm p}(t) \right) \\
		B^{\rm in}_{{\bm p}}(t) =  {\rm e}^{+\frac{{\rm i}\pi}{8}} {\rm e}^{-\frac{\pi a_{\bm p}}{4}} D_{{\rm i} a_{\bm p} } \left(-{\rm e}^{-\frac{{\rm i}\pi}{4}}\xi_{\bm p}(t) \right)
	\end{array}\right. \;, \nonumber\\
	&\left\{\begin{array}{l}
		A^{\rm out}_{{\bm p}}(t) =  {\rm e}^{-\frac{{\rm i}\pi}{8}} {\rm e}^{-\frac{\pi a_{\bm p}}{4}}  D_{-{\rm i} a_{\bm p} } \left({\rm e}^{\frac{{\rm i}\pi}{4}} \xi_{\bm p}(t) \right) \\
		B^{\rm out}_{{\bm p}}(t) =  {\rm e}^{+\frac{{\rm i}\pi}{8}} {\rm e}^{-\frac{\pi a_{\bm p}}{4}} \sqrt{a_{\bm p}}  D_{-{\rm i} a_{\bm p} -1} \left({\rm e}^{\frac{{\rm i}\pi}{4}}\xi_{\bm p}(t) \right)
	\end{array}\right. \;,  \label{eq----8}
\end{align}
where $D_{\nu}(z)$ is the parabolic cylinder function and  
\begin{align}
	a_{\bm p} := \frac{m^2 + p_{\perp}^2}{2e\bar{E}} \ \ {\rm and}\ \ 
	\xi_{\bm p}(t) := \sqrt{\frac{2}{e\bar{E}}}(e\bar{E}t + p_z) \;. 
\end{align}
The four component spinors $\Gamma_s$, with $s = \pm 1$ specifying the spin direction with respect to ${\bm e}_z$, are defined as the two eigenvectors of $\gamma^0 \gamma^3$ with the eigenvalue one; e.g., in the Dirac representation, $\Gamma_s$ read
\begin{align}
	\Gamma_{+1} := \frac{1}{\sqrt{2}} \begin{pmatrix} 1 \\ 0 \\ 1 \\ 0 \end{pmatrix} \;, \ 
	\Gamma_{-1} := \frac{1}{\sqrt{2}} \begin{pmatrix} 0 \\ 1 \\ 0 \\ -1 \end{pmatrix} \;.  
\end{align}

For $\bar{E} = 0$, the spinor decomposition (\ref{eqd31}) needs not to be modified, but the scalar functions $A^{\rm as}_{{\bm p}}$ and $B^{\rm as}_{{\bm p}}$ change.  Substituting the spinor decomposition (\ref{eqd31}) into the free Dirac equation, we immediately find that $A^{\rm as}_{{\bm p}}$ and $B^{\rm as}_{{\bm p}}$ satisfy the usual free Klein-Gordon equation and, after imposing the boundary conditions (\ref{eq-10}) and (\ref{aaaeq-10}), get
\begin{align}
	&\left\{\begin{array}{l}
		A^{\rm in}_{{\bm p}} = A^{\rm out}_{{\bm p}}
		= \frac{1}{\sqrt{2}}\sqrt{1+\frac{p_z}{\sqrt{m^2+{\bm p}^2}}} {\rm e}^{-{\rm i}\sqrt{m^2+{\bm p}^2}t} \\
		B^{\rm in}_{{\bm p}} = B^{\rm out}_{{\bm p}} 
		= \frac{1}{\sqrt{2}}\sqrt{1-\frac{p_z}{\sqrt{m^2+{\bm p}^2}}} {\rm e}^{-{\rm i}\sqrt{m^2+{\bm p}^2}t}
	\end{array}\right. \;.  \label{eqa---8}
\end{align}

Below, we focus on the case of $\bar{E} \neq 0$.  The $\bar{E} = 0$ case can be done in the same manner just by replacing the scalar functions for $\bar{E} \neq 0$ (\ref{eq----8}) to those for $\bar{E} = 0$ (\ref{eqa---8}).  Or, the $\bar{E} = 0$ results should be able to be obtained by taking the $\bar{E} \to 0$ limit of the $\bar{E} \neq 0$ results.

\subsection{Imaginary part ${\rm Im}\,\epsilon_{ij}$} \label{app:A}

From Eq.~(\ref{erqasdasda:36}), what we need to evaluate is the Fourier transformation of $\Pi_{ij}$, i.e., 
\begin{align}
	\tilde{\Pi}_{ij} (\omega)
	&= 2e^2  \int^{+\infty}_{-\infty}{\rm d}\tau\,{\rm e}^{+{\rm i}\omega \tau}\,{\rm Im}\,{\rm tr}  \sum_{s,s'} \int \frac{{\rm d}^3{\bm p}}{(2\pi)^3} \nonumber\\
	&\quad \times \gamma^i  S^{\rm in}_{-,{\bm p},s}\left( +\frac{\tau}{2}, -\frac{\tau}{2} \right)  \gamma^j   S^{\rm in}_{+,{\bm p},s'} \left( -\frac{\tau}{2},+\frac{\tau}{2} \right) \nonumber\\
	&= -{\rm i}e^2 \sum_{s,s'} \int \frac{{\rm d}^3{\bm P}}{(2\pi)^3} \int^{+\infty}_{-\infty}\frac{{\rm d}k}{2\pi} \nonumber\\
		&\quad \times \left[ \left[ \tilde{\Gamma}^i_{{\bm P},s,s'}\left(k\right) \right]^\dagger \tilde{\Gamma}^j_{{\bm P},s,s'}\left(-\omega - k\right) \right. \nonumber\\
		&\quad\quad\quad \left. -   \left[ \tilde{\Gamma}^j_{{\bm P},s,s'}\left(+\omega-k\right) \right]^\dagger \tilde{\Gamma}^i_{{\bm P},s,s'}\left(k\right) \right] \;, \label{qerqer:a5g}
\end{align}
where
\begin{align}
	&\tilde{\Gamma}^i_{{\bm P},s,s'}\left( \omega \right) \nonumber\\
	&:= \int^{+\infty}_{-\infty}{\rm d}\tau\,{\rm e}^{+{\rm i}\omega \tau}\,\bar{\Psi}^{\rm in}_{-,{\bm P},s} \left(\frac{\tau}{2}\right) \gamma^i \Psi^{\rm in}_{+,{\bm P},s'}\left(\frac{\tau}{2}\right) \;. 
\end{align}
We first use Eq.~(\ref{tqewda:31}) and factor out the dependence on $P_z$ of $\tilde{\Gamma}^i_{{\bm P},s,s'}$ as
\begin{align}
   &\tilde{\Gamma}^i_{{\bm P},s,s'}\left(\omega\right) \nonumber\\
	&= \sqrt{\frac{2}{e\bar{E}}} {\rm e}^{-2{\rm i} \frac{\omega P_z}{e\bar{E}}} \nonumber\\
		&\quad \times \int^{+\infty}_{-\infty}{\rm d}\xi \,{\rm e}^{+{\rm i} \omega \sqrt{\frac{2}{e\bar{E}}} \xi } \bar{\psi}^{\rm in}_{-,{\bm P}-P_z{\bf e}_z,s} \left( \xi \right) \gamma^i  \psi^{\rm in}_{+,{\bm P}-P_z{\bf e}_z,s'}\left( \xi \right) \nonumber\\
	&=: \sqrt{\frac{2}{e\bar{E}}} {\rm e}^{-2{\rm i} \frac{\omega P_z}{e\bar{E}} }\times \tilde{\Gamma}^{\prime i}_{{\bm p}_\perp,s,s'}\left( \omega \sqrt{\frac{2}{e\bar{E}}} \right) \;, \label{t23r3232:a8}
\end{align}
where we changed the integration variable $\tau$ to $\xi := \xi_{\bm P}(\tau/2)$.  The function $\tilde{\Gamma}^{\prime i}_{{\bm p}_\perp,s,s'}$ is independent of $P_z$.  We can then carry out the $P_z$ integration in Eq.~(\ref{qerqer:a5g}) to obtain
\begin{align}
	&\tilde{\Pi}_{ij} (\omega)\\
	&= -2 {\rm i} \alpha \sum_{s,s'} \int \frac{{\rm d}^2{\bm p}_\perp}{(2\pi)^2} \nonumber\\
		&\quad \times \left[ \left[ \tilde{\Gamma}^{\prime i}_{{\bm p}_\perp,s,s'}\left(-\frac{\omega}{\sqrt{2e\bar{E}}}\right) \right]^\dagger \tilde{\Gamma}^{\prime j}_{{\bm p}_\perp,s,s'}\left(-\frac{\omega}{\sqrt{2e\bar{E}}}\right)  \right. \nonumber\\
		&\quad\quad \left.  -   \left[  \tilde{\Gamma}^{\prime j}_{{\bm p}_\perp,s,s'}\left(+\frac{\omega}{\sqrt{2e\bar{E}}}\right) \right]^\dagger \tilde{\Gamma}^{\prime i}_{{\bm p}_\perp,s,s'}\left(+\frac{\omega}{\sqrt{2e\bar{E}}}\right) \right] \;. \nonumber
\end{align}

Thus, our central task is to evaluate $\tilde{\Gamma}^{\prime i}_{{\bm p}_\perp,s,s'}$ (\ref{t23r3232:a8}).  Using the solutions of the Dirac equation (\ref{eqd31}) and by simplifying the spinor structure, we find
\begin{subequations}
\begin{align}
	&\left[ \tilde{\Gamma}^{\prime 1}_{{\bm p}_\perp,s',s}\left( \omega \right) \right]^\dagger \\
	&= - \left[ \frac{p_x + {\rm i}sp_y }{m_\perp}  I[A^{{\rm in}*}_{{\bm p}}A^{{\rm in}*}_{{\bm p}};\omega] \right. \nonumber\\
		&\quad\quad\quad \left. - \frac{p_x - {\rm i}sp_y }{m_\perp}  I[B^{{\rm in}*}_{{\bm p}}B^{{\rm in}*}_{{\bm p}};\omega] \right] \delta_{s,s'}  \nonumber\\
		&\quad - s \frac{m}{m_\perp} \left[ I[A^{{\rm in}*}_{{\bm p}}A^{{\rm in}*}_{{\bm p}};\omega] + I[B^{{\rm in}*}_{{\bm p}}B^{{\rm in}*}_{{\bm p}};\omega] \right] \delta_{s,-s'} \;, \nonumber\\
	&\left[ \tilde{\Gamma}^{\prime 2}_{{\bm p}_\perp,s',s}\left( \omega \right) \right]^\dagger \\
	&= +{\rm i}s \left[ \frac{p_x + {\rm i}sp_y }{m_\perp}  I[A^{{\rm in}*}_{{\bm p}}A^{{\rm in}*}_{{\bm p}};\omega] \right. \nonumber\\
		&\quad\quad\quad \left. +  \frac{p_x - {\rm i}sp_y }{m_\perp}  I[B^{{\rm in}*}_{{\bm p}}B^{{\rm in}*}_{{\bm p}};\omega] \right] \delta_{s,s'} \nonumber\\
		&\quad + {\rm i} \frac{m}{m_\perp} \left[ I[A^{{\rm in}*}_{{\bm p}}A^{{\rm in}*}_{{\bm p}};\omega] + I[B^{{\rm in}*}_{{\bm p}}B^{{\rm in}*}_{{\bm p}};\omega] \right] \delta_{s,-s'} \;, \nonumber \\
	&\left[ \tilde{\Gamma}^{\prime 3}_{{\bm p}_\perp,s',s}\left( \omega \right) \right]^\dagger 
	= - 2 I[A^{{\rm in}*}_{{\bm p}}B^{{\rm in}*}_{{\bm p}};\omega]  \delta_{s,s'}  \;,
\end{align}
\end{subequations}
where  
\begin{align}
	I[XY;\omega] 
	:= \int {\rm d}\xi\,{\rm e}^{+{\rm i}\omega \xi} X(\xi) Y(\xi) \;. \label{trqrd:a11}
\end{align}
Using Eq.~(\ref{eq----8}), the integrals $I[A^{{\rm in}*}_{{\bm p}}A^{{\rm in}*}_{{\bm p}}], I[A^{{\rm in}*}_{{\bm p}}B^{{\rm in}*}_{{\bm p}}]$, and $I[B^{{\rm in}*}_{{\bm p}}B^{{\rm in}*}_{{\bm p}}]$ can be expressed as 
\begin{subequations}
\begin{align}
	I[A^{{\rm in}*}_{{\bm p}}B^{{\rm in}*}_{{\bm p}}; \omega]
		&= {\rm e}^{- \frac{\pi a_{\bm p}}{2}} \sqrt{a_{\bm p}} {\mathcal I}_{1,0}(\omega) \; , \\[12pt]
	I[B^{{\rm in}*}_{{\bm p}}B^{{\rm in}*}_{{\bm p}};\omega]
		&= {\rm e}^{- \frac{\pi a_{\bm p}}{2}} {\rm e}^{-{\rm i}\pi/4} {\mathcal I}_{0,0}(\omega)  \; ,  \\[12pt]
	I[A^{{\rm in}*}_{{\bm p}}A^{{\rm in}*}_{{\bm p}};\omega]
		&= {\rm e}^{- \frac{\pi a_{\bm p}}{2}} a_{\bm p} {\rm e}^{+{\rm i}\pi/4} {\mathcal I}_{1,1}(\omega) \; ,
\end{align}
\end{subequations}
where
\begin{align}
	&{\mathcal I}_{\lambda,\lambda'}(\omega) \label{gqefrqerqacp} \\
	&:= \int {\rm d}\xi\, {\rm e}^{+{\rm i} \omega \xi}  D_{-{\rm i} a_{\bm p} - \lambda } \left(-{\rm e}^{+\frac{{\rm i}\pi}{4}} \xi \right) D_{-{\rm i} a_{\bm p} - \lambda' } \left(-{\rm e}^{+\frac{{\rm i}\pi}{4}} \xi \right) \; . \nonumber
\end{align}
The evaluation of the integral ${\mathcal I}_{\lambda,\lambda'}$ is a bit tedious but is analytically doable.  Using the integral representation of the parabolic cylinder function~\cite{Bateman:100233}, 
\begin{align}
	&D_{\nu}(-{\rm e}^{+{\rm i}\pi/4} \xi) \nonumber\\
	&= \frac{{\rm e}^{-{\rm i}\xi^2/4}}{{\rm e}^{+{\rm i}\pi\nu/4}\Gamma(-\nu)} \int_0^{\infty} {\rm d}y \, y^{-\nu-1}{\rm e}^{+{\rm i} \xi y - {\rm i} y^2/2} \; ,   \label{gr2eq:a14}
\end{align}
we can exactly carry out the $\xi$ integration in ${\mathcal I}_{\lambda,\lambda'}$ (\ref{gqefrqerqacp}).  The remaining two $y$ integrations can also be done, and the result can be expressed in terms of the Kummer's and Tricomi's confluent hypergeometric functions, denoted by ${}_1 F_1(a,b;z)$ and $U(a,b;z) $, respectively:
\begin{align}
	&{\mathcal I}_{\lambda,\lambda'}(\omega) \label{gqerqerq:a15} \\
	&= \sqrt{2\pi} {\rm e}^{+{\rm i}\pi(\lambda-\lambda'-1)/4} |\omega|^{\lambda-\lambda'} \Gamma(-{\rm i}a_{\bm p}-\lambda'+1) \nonumber\\
		&\quad \times \left[ \Theta(-\omega) \left\{ \frac{ {\rm e}^{-\pi a_{\bm p}} {\rm e}^{{\rm i}\pi\lambda'} }{ \Gamma({\rm i}a_{\bm p}+\lambda) } \right.\right. \nonumber\\
			&\quad\quad\quad\quad\quad\quad\quad \times  {\rm e}^{-\frac{{\rm i}\omega^2}{2}}  U\left( -{\rm i}a_{\bm p}-\lambda'+1; \lambda-\lambda'+1; +{\rm i}\omega^2  \right)	 \nonumber\\
		&\quad\quad\quad\quad\quad\quad  +  \frac{1}{ \Gamma(\lambda-\lambda'+1) } {\rm e}^{+\frac{{\rm i}\omega^2}{2}} \nonumber\\
			&\quad\quad\quad\quad\quad\quad\quad \left.  \times  {}_1\tilde{F}_1\left( {\rm i}a_{\bm p}+\lambda; \lambda-\lambda'+1; -{\rm i}\omega^2 \right) \right\} \nonumber\\
		&\quad\quad \left. + \Theta(+\omega) \frac{ {\rm e}^{-\pi a_{\bm p}} {\rm e}^{{\rm i}\pi\lambda'} }{  \Gamma(-{\rm i}a_{\bm p}-\lambda'+1) } \right. \nonumber\\
		&\quad\quad\quad\quad\quad\quad\quad \times \left. {\rm e}^{+\frac{{\rm i}\omega^2}{2}} U\left( {\rm i}a_{\bm p}+\lambda; \lambda-\lambda'+1; - {\rm i}\omega^2 \right) \right] \; . \nonumber
\end{align}

Now, all the integrations [except for the transverse-momentum $p_\perp$ integration in $\tilde{\Pi}_{ij}$ (\ref{qerqer:a5g}), which seems infeasible analytically] have been done.  The remaining task is quite straightforward (though quite bothersome): just substitute the obtained ${\mathcal I}_{\lambda,\lambda'}$ (\ref{gqerqerq:a15}) into $\tilde{\Pi}_{ij}$ (\ref{qerqer:a5g}) and simplify it using identities of the confluent hypergeometric functions~\cite{Bateman:100233}.  Doing this yields Eq.~(\ref{g2e;::15}) in the main text.

\subsection{Number of pairs $N$} \label{app:B}

The evaluation of the number of pairs $N$ (\ref{eq:13}) can be done in a similar manner to ${\rm Im}\,\epsilon_{ij}$ (see Sec.~\ref{app:A}).  We just have to plug the mode functions (\ref{eqd31}) into Eq.~(\ref{eq:13}) and then calculate the corresponding Fourier transformations of the products among the scalar functions $A^{\rm as}_{\bm p}$ and $B^{\rm as}_{\bm p}$.

To be concrete, we substitute Eq.~(\ref{eqd31}) into Eq.~(\ref{eq:13}).  The first term in the brackets can be evaluated to~\cite{Nikishov:1969tt, Tanji:2008ku}
\begin{align}
	\psi^{{\rm out}\dagger}_{+,{\bm p},s} \psi^{\rm in}_{-,{\bm p},s'} 
	= {\rm e}^{-\pi a_{\bm p}} \delta_{s,s'} \;.
\end{align}
The remaining second term gives
\begin{align}
	&\int {\rm d}t \, \bar{\psi}^{\rm out}_{+,{\bm p},s} \slashed{\mathcal A} \psi^{\rm in}_{-,{\bm p},s'} \\
	&= \delta_{s,s'} \int \frac{{\rm d}\omega}{2\pi} \frac{ {\rm e}^{-{\rm i}\frac{\omega p_z}{e\bar{E}}} }{\sqrt{2e\bar{E}}} \nonumber\\
		&\quad \times \left[ (\tilde{\mathcal A}^1(-\omega)-{\rm i}s\tilde{\mathcal A}^2(-\omega)) \right. \nonumber\\ 
				&\quad\quad\quad\quad \times \frac{p_x+{\rm i}sp_y}{\sqrt{m^2+{\bm p}_\perp^2}} I\left[ A^{{\rm in}*}_{{\bm p}} A^{{\rm out}*}_{{\bm p}} ; \frac{\omega}{\sqrt{2e\bar{E}}} \right] \nonumber\\
			&\quad\quad  - (\tilde{\mathcal A}^1(-\omega)+{\rm i}s\tilde{\mathcal A}^2(-\omega))  \nonumber\\ 
				&\quad\quad\quad\quad \times \frac{p_x-{\rm i}sp_y}{\sqrt{m^2+{\bm p}_\perp^2}} I\left[ B^{{\rm in}*}_{{\bm p}} B^{{\rm out}*}_{{\bm p}} ; \frac{\omega}{\sqrt{2e\bar{E}}} \right] \nonumber\\
			&\quad\quad  - \tilde{\mathcal A}^3(-\omega)\left(   I\left[ A^{{\rm in}*}_{{\bm p}} B^{{\rm out}*}_{{\bm p}} ; \frac{\omega}{\sqrt{2e\bar{E}}} \right]   \right. \nonumber\\ 
				&\quad\quad\quad\quad\quad\quad\quad \left. \left. +  I\left[ B^{{\rm in}*}_{{\bm p}} A^{{\rm out}*}_{{\bm p}} ; \frac{\omega}{\sqrt{2e\bar{E}}} \right] \right) \right]   \nonumber\\
		&\quad  +  (1-\delta_{s,s'})s\int \frac{{\rm d}\omega}{2\pi} \frac{ {\rm e}^{-{\rm i}\frac{\omega p_z}{e\bar{E}}} }{\sqrt{2e\bar{E}}} \nonumber\\
			&\quad\quad \times (\tilde{\mathcal A}^1(-\omega)-{\rm i}s\tilde{\mathcal A}^2(-\omega)) \frac{m}{\sqrt{m^2+{\bm p}_\perp^2}} \nonumber\\
			&\quad\quad \times  \left(  I\left[ A^{{\rm in}*}_{{\bm p}} A^{{\rm out}*}_{{\bm p}} ; \frac{\omega}{\sqrt{2e\bar{E}}} \right]  +   I\left[ B^{{\rm in}*}_{{\bm p}} B^{{\rm out}*}_{{\bm p}} ; \frac{\omega}{\sqrt{2e\bar{E}}} \right]  \right) \;. \nonumber
\end{align}
The functions $I[\cdots]$ are defined in the same manner as Eq.~(\ref{trqrd:a11}).  Using the integral representation of the parabolic cylinder function (\ref{gr2eq:a14}), it is possible to exactly carry out the $\xi$ integration in $I$, which yields~\cite{Huang:2019uhf}
\begin{subequations}
\begin{align}
	&I\left[ A^{{\rm in}*}_{{\bm p}} A^{{\rm out}*}_{{\bm p}} ; \frac{\omega}{\sqrt{2e\bar{E}}} \right] \nonumber\\
	&= 2\pi {\rm e}^{{\rm i}\pi/4} \Theta(\omega)  \sqrt{a_{\bm p}}  {\rm e}^{-\pi a_{\bm p}}  \frac{\sqrt{2e\bar{E}} }{\omega} M_{\frac{1}{2}+{\rm i}a_{\bm p}, 0}\left( {\rm i} \frac{\omega^2}{2e\bar{E}} \right)  \;, \\
	&I\left[ A^{{\rm in}*}_{{\bm p}} B^{{\rm out}*}_{{\bm p}} ; \frac{\omega}{\sqrt{2e\bar{E}}} \right] = I\left[ B^{{\rm in}*}_{{\bm p}} A^{{\rm out}*}_{{\bm p}} ; \frac{\omega}{\sqrt{2e\bar{E}}} \right] \nonumber\\
	&= 2\pi {\rm i} \Theta(\omega)   a_{\bm p} {\rm e}^{-\pi a_{\bm p}}  \frac{\sqrt{2e\bar{E}} }{\omega} M_{{\rm i}a_{\bm p}, \frac{1}{2}}\left( {\rm i} \frac{\omega^2}{2e\bar{E}} \right) \;, \\
	&I\left[ B^{{\rm in}*}_{{\bm p}} B^{{\rm out}*}_{{\bm p}} ; \frac{\omega}{\sqrt{2e\bar{E}}} \right] \nonumber\\
	&= 2\pi {\rm e}^{-{\rm i}\pi/4} \Theta(\omega)  \sqrt{a_{\bm p}}  {\rm e}^{-\pi a_{\bm p}}  \frac{\sqrt{2e\bar{E}} }{\omega} M_{\frac{1}{2}-{\rm i}a_{\bm p}, 0}\left( -{\rm i} \frac{\omega^2}{2e\bar{E}} \right) \;.
\end{align}
\end{subequations}
Note that the Whittaker function $M_{\kappa, \mu}(z)$ is related to the Kummer's confluent hypergeometric function ${}_1F_1(a,b;z)$ as $M_{\kappa, \mu}(z) = \exp[-z/2] z^{\mu+1/2} {}_1F_1(\mu-\kappa+1/2,1+2\mu;z)$.

Putting everything together and carrying out the $p_z$ integration in Eq.~(\ref{eq:13}), we obtain
\begin{align}
	N
	&= \frac{m^2}{2\pi^2} \int_0^\infty {\rm d}p_\perp\, p_\perp  {\rm e}^{-2\pi a_{\bm p}}  \nonumber\\
		&\quad \times \Bigg[ V_4 \frac{eE}{m^2}  \nonumber\\
		&\quad\quad + e^2 \frac{4\pi^2}{m^2} V_3 \int_0^\infty \frac{{\rm d}\omega}{2\pi}  \frac{m_\perp^2}{\omega^2} (|\tilde{\mathcal A}^1(\omega)|^2 + |\tilde{\mathcal A}^2(\omega)|^2) \nonumber\\
			&\quad\quad\quad\quad \times  \left| M_{\frac{1}{2}+{\rm i}a_{\bm p}, 0}\left( {\rm i} \frac{\omega^2}{2eE} \right) \right|^2   \nonumber\\
		&\quad\quad + e^2 \frac{4\pi^2}{m^2} V_3 \int_0^\infty \frac{{\rm d}\omega}{2\pi} \frac{m_\perp^2}{eE} \frac{m_\perp^2}{\omega^2}  |\tilde{\mathcal A}^3(\omega)|^2 \left|  M_{{\rm i}a_{\bm p}, \frac{1}{2}}\left( {\rm i} \frac{\omega^2}{2eE} \right)  \right|^2    \nonumber\\
		&\quad\quad  - e^2 \frac{4\pi^2}{m^2} V_3 \int_0^\infty \frac{{\rm d}\omega}{2\pi} \frac{m^2}{\omega^2}  (|\tilde{\mathcal A}^1(\omega)|^2  +  |\tilde{\mathcal A}^2(\omega)|^2  ) \nonumber\\
			&\quad\quad\quad\quad \times  {\rm Im}\left[ \left\{ M_{\frac{1}{2}+{\rm i}a_{\bm p}, 0}\left( {\rm i} \frac{\omega^2}{2eE} \right)  \right\}^2 \right] \Bigg] \;. \label{eq::::::::::::b47}
\end{align}
Note that we have assumed $\omega \neq 0$ and used an identity $\int {\rm d}p_z = e\bar{E} V_4/V_3$~\cite{Nikishov:1969tt}.  Equation~(\ref{gqerqewqrwq:60}) in the main text is for the setup (\ref{eq:2}), i.e., 
\begin{align}
	\tilde{\bm {\mathcal A}} = \frac{{\bm {\mathcal E}}}{\omega} \pi \left[ {\rm e}^{+{\rm i}\phi}\delta(\omega-\omega)  +  {\rm e}^{-{\rm i}\phi}\delta(\omega+\omega)   \right] \;.
\end{align}

\section{Heisenberg-Euler analysis} \label{app:C}

The Heisenberg-Euler effective Lagrangian is the one-loop effective action of QED in the presence of constant electric ${\bm E}$ and magnetic ${\bm B}$ fields~\cite{Heisenberg:1936nmg}: 
\begin{align}
	{\mathcal L} 
	&= -{\mathcal F} - \frac{1}{8\pi^2} \int_0^\infty \frac{{\rm d}s}{s^3} {\rm e}^{-m^2s}  \label{gqwereqwdwq:c2} \\
		&\quad \times \left[ (es)^2 \frac{{\rm Re}\,\cosh\left(  es \sqrt{2({\mathcal F}+{\rm i}{\mathcal G})}  \right)}{ {\rm Im}\,\cosh\left(  es \sqrt{2({\mathcal F}+{\rm i}{\mathcal G})}  \right) }{\mathcal G}  -   \frac{2}{3}(es)^2{\mathcal F}  -   1   \right] \;,  \nonumber
\end{align}
where ${\mathcal F} := ( {\bm B}^2 - {\bm E}^2 )/2$ and ${\mathcal G} := {\bm E} \cdot {\bm B}$.  For the purely electric case ${\bm B} = {\bm 0}$, Eq.~(\ref{gqwereqwdwq:c2}) can be simplified to 
\begin{align}
	{\mathcal L} 
	&= \frac{1}{2}E^2 - \frac{1}{8\pi^2} (eE)^2 \int_0^{\infty} \frac{{\rm d}y}{y^3} {\rm e}^{ - \frac{m^2}{eE} y } \left[ \frac{ y }{  \tan y }   +   \frac{y^2}{3}  -   1   \right] \;,  \label{eq::f;2r1} 
\end{align}
where $y := eEs $.  Assuming the electric field is weak, it is possible to (formally) expand Eq.~(\ref{eq::f;2r1}) in terms of the field strength $eE/m^2$.  Expanding the terms in the square brackets in $y$ and then performing the $y$ integration, 
\begin{align}
	{\mathcal L} 
	&= \frac{1}{2}E^2 - \frac{m^4}{8\pi^2} \sum_{\ell = 0}^\infty \frac{(-4)^{\ell+2}  (2\ell+1)! }{(2\ell+4)!} B_{2\ell+4}  \left( \frac{eE}{m^2} \right)^{2\ell+4} \nonumber\\
	&= \frac{1}{2}E^2  +  \frac{m^4}{8\pi^2} \left[ \frac{1}{45} \left( \frac{eE}{m^2} \right)^{4}  +   \frac{4}{315} \left( \frac{eE}{m^2} \right)^{6}  + \cdots  \right] \;, \label{gweqfdqed:c3}
\end{align}
where $B_n$ is the Bernoulli number.  Note that Eq.~(\ref{gweqfdqed:c3}) does not contain any non-perturbative factors ${\rm e}^{-n\pi m^2/eE}$ ($n \in {\mathbb N}$), which are implicitly dropped when doing the expansion.  Due to the absence of the non-perturbative factor, the Lagrangian (\ref{gweqfdqed:c3}) is purely real, i.e., there is no dynamical process (i.e., the Schwinger effect) and the system is implicitly assumed to be in equilibrium.

We are interested in a situation where the constant electric field ${\bm E}$ can be divided into two as 
\begin{align}
	{\bm E} = \bar{\bm E} + {\bm {\mathcal E}} \;,
\end{align} 
which corresponds to $\omega \to 0$ limit of the field configuration in the main text (\ref{eq:2}).  The electric displacement vector for the probe electric field ${\bm {\mathcal E}}$ is given by~\cite{Heisenberg:1936nmg, lan84}
\begin{align}
	{\bm {\mathcal D}} = \frac{\delta {\mathcal L}}{\delta {\bm {\mathcal E}} } \;. \label{gqerqer:c3}
\end{align}
We emphasize that it is implicitly assumed here that the electric field must be weak.  For a strong electric field, the Heisenberg-Euler effective Lagrangian acquires a non-zero imaginary part $\propto {\rm e}^{-\pi m^2/eE}$~\cite{Heisenberg:1936nmg, Schwinger:1951nm}.  If so, i.e., the non-perturbative factor is non-negligible ${\rm e}^{-\pi m^2/eE} \approx \!\!\!\!\!\!/ \; 0$, the corresponding electric displacement vector (\ref{gqerqer:c3}) becomes complex, which is not allowed by the definition.  Note also that Eq.~(\ref{gqerqer:c3}) is consistent with our identification of the displacement vector (\ref{gqerqed:6}).  Indeed, Eq.~(\ref{gqerqer:c3}) indicates $\partial_t (\delta {\mathcal L}/\delta \dot{\bm {\mathcal A}} ) = -\dot{\bm {\mathcal D}}$.  Also, the external current ${\bm J}_{\rm ext}$ is related to the effective Lagrangian ${\mathcal L}$ as ${\bm J}_{\rm ext} = \delta {\mathcal L}/\delta {\bm {\mathcal A}} $ because of the gauge invariance.  Combining those two, we understand that the Euler-Lagrange equation gives our identification (\ref{gqerqed:6}).

The electric permittivity can now be obtained straightforwardly from the standard definition (\ref{fqereqr;q}).  Now that the electric field and the displacement vectors are constants in the coordinate space, i.e., ${\bm {\mathcal X}}(t) = {\bm {\mathcal X}}$ (${\bm {\mathcal X}} = {\bm {\mathcal E}}$ and ${\bm {\mathcal D}}$), their Fourier components are simply given by $\tilde{\bm {\mathcal X}}(\omega) = {\bm {\mathcal X}} \delta(\omega)$ and hence $\delta \tilde{{\bm {\mathcal D}}}/\delta \tilde{\bm {\mathcal E}} = \delta {\bm {\mathcal D}}/\delta {\bm {\mathcal E}}$ holds.  Then, from Eqs.~(\ref{fqereqr;q}) and (\ref{gqerqer:c3}), we obtain
\begin{align}
	\epsilon_{ij}
	= \left. \frac{\delta \tilde{\mathcal D}_i}{\delta \tilde{\mathcal E}_j} \right|_{{\bm {\mathcal E}} \to {\bm 0}}
	= \left. \frac{\delta {\mathcal D}_i}{\delta {\mathcal E}_j} \right|_{{\bm {\mathcal E}} \to {\bm 0}}
	= \left. \frac{\delta^2 {\mathcal L}}{\delta {\mathcal E}_i \delta {\mathcal E}_j} \right|_{{\bm {\mathcal E}} \to {\bm 0}} \;. \label{qerqedwq:c6}
\end{align}

Substituting the weak-field expansion (\ref{gweqfdqed:c3}) into the electric permittivity (\ref{qerqedwq:c6}), we get
\begin{align}
	\epsilon_{ij}
	&= \delta_{ij}  
		-  \frac{\alpha}{2\pi} \sum_{\ell = 0}^\infty \left( \frac{e\bar{E}}{m^2} \right)^{2\ell+2}  B_{2\ell+4}  \frac{ (-4)^{\ell+2} }{(2\ell+3)(2\ell+2)} \nonumber\\
		&\quad \times \left( \delta_{ij}   +   (2\ell + 2) \frac{\bar{E}_i \bar{E}_j}{\bar{E}^2} \right) \nonumber\\
	&= \delta_{ij}  
		+  \frac{\alpha}{2\pi} \left[ \frac{4}{45} \left( \delta_{ij} + 2 \frac{\bar{E}_i \bar{E}_j}{\bar{E}^2} \right) \left( \frac{e\bar{E}}{m^2} \right)^{2}    \right. \nonumber\\
		&\quad \left. +  \frac{8}{105} \left( \delta_{ij}   +   4 \frac{\bar{E}_i \bar{E}_j}{\bar{E}^2} \right) \left( \frac{e\bar{E}}{m^2} \right)^{4}  \cdots \right] \;.
\end{align}
The lowest-order expression was derived independently by Klein-Nigam~\cite{PhysRev.135.B1279} and Baier-Breitenlohner~\cite{Baier:1967zzc} in the 1960s.  Notice that the electric permittivity is purely real, i.e.,  
\begin{align}
	{\rm Im}\,\epsilon_{ij} = 0 \;,
\end{align}
because, as we have stressed, the weak-field expansion is assumed implicitly in the Heisenberg-Euler approach.  Setting ${\bm E} = \bar{E} {\bf e}_z$ and subtracting the $\bar{E}=0$ contribution, we find that the change $\Delta \epsilon_{ij}$ is given by
\begin{subequations}
\begin{align}
	\Delta \epsilon_{\perp}
	&= {\rm Re}\,\Delta \epsilon_{\perp} \nonumber\\
	&= - \frac{\alpha}{2\pi} \sum_{\ell = 0}^\infty \left( \frac{e\bar{E}}{m^2} \right)^{2\ell+2}  B_{2\ell+4}  \frac{ (-4)^{\ell+2} }{2\ell+2} \frac{1}{2\ell+3}  \nonumber\\
	&= +  \frac{\alpha}{\pi} \left[ \frac{2}{45}  \left( \frac{e\bar{E}}{m^2} \right)^{2}    +     \frac{4}{105} \left( \frac{e\bar{E}}{m^2} \right)^{4}    +    \cdots \right] \;. \\
	\Delta \epsilon_{\parallel} 
	&= 	{\rm Re}\,\Delta \epsilon_{\parallel} \nonumber\\
	&= -  \frac{\alpha}{2\pi} \sum_{\ell = 0}^\infty \left( \frac{e\bar{E}}{m^2} \right)^{2\ell+2}  B_{2\ell+4}  \frac{ (-4)^{\ell+2} }{2\ell+2} \nonumber\\
	&= +  \frac{\alpha}{\pi} \left[ 3 \frac{2}{45} \left( \frac{e\bar{E}}{m^2} \right)^{2}    +    5 \frac{4}{105}  \left( \frac{e\bar{E}}{m^2} \right)^{4}   +    \cdots \right] \;.
\end{align}
\end{subequations}
The lowest-order terms correspond to Eq.~(\ref{fqefedada:74}) in the main text.

\bibliography{bib}
\end{document}